\documentclass{article}
\usepackage{graphicx} % Required for inserting images
\usepackage{float}  
\usepackage{authblk}
\usepackage{xcolor}
\usepackage{amsmath} 
\usepackage{soul}

\usepackage{blindtext}
\usepackage[a4paper, total={6in, 8in}]{geometry}
\usepackage[
  backend=biber,
  style=numeric,
  sorting=none
]{biblatex}
\title{
Collective synchrony in confluent, pulsatile epithelia}
\author[1]{Wenhui Tang$^*$}
\author[2,3]{Mehrana R. Nejad$^*$}
\author[4]{Adrian F. Pegoraro}
\author[2,3]{L. Mahadevan$^\dag$}
\author[1]{Ming Guo$^\dag$}
\affil[1]{Department of Mechanical Engineering, Massachusetts Institute of Technology, Cambridge, Massachusetts, USA}
\affil[2]{John A. Paulson School of Engineering and Applied Sciences, Harvard University,
Cambridge, Massachusetts, USA}
\affil[3]{Departments of Physics and Organismic and Evolutionary Biology, Harvard University, Cambridge, Massachusetts, USA}
\affil[4]{Metrology Research Centre, National Research Council Canada, Ottawa, Ontario, Canada}
\date{}
\begin{document}

\maketitle
\def\thefootnote{*}\footnotetext{These authors contributed equally to this work.}\def\thefootnote{\arabic{footnote}}
\def\thefootnote{\dag}\footnotetext{Corresponding authors: lmahadev@g.harvard.edu; guom@mit.edu}\def\thefootnote{\arabic{footnote}}
\section*{Abstract}
Collective cell migration lies at the intersection of developmental biology and non-equilibrium physics, where active processes give rise to emergent patterns that are biologically relevant. Here, we investigate dilatational modes—cycles of expansion and contraction—in epithelial monolayers, and show that the divergence of the velocity field exhibits robust, large-scale temporal oscillations. These oscillatory patterns, reminiscent of excitable media and their biological analogs, emerge spontaneously from the coupled dynamics of actively pulsing cells. We find that the temporal persistence of these oscillations varies non-monotonically with cell density: synchrony initially increases with density, reaches a maximum at intermediate densities and is lost at higher values. This trend mirrors changes in the spatial correlation length of cell-cell interactions, and the density of topological defects in the system, suggesting a shared physical origin. We develop a continuum model in which a complex-valued Ginzburg–Landau-type field that governs the amplitude and phase of oscillations is coupled to local cell density. Simulations reproduce the observed behavior, revealing that local density adapts to phase patterns, reinforcing temporal coherence up to a critical density, and variations in the density of topological defects as a function of cell density. Extending our analysis to breast cancer cell lines with increasing invasiveness, we find that malignant cells exhibit longer phase persistence and fewer topological defects, suggesting a mechanistic link between temporal coherence and metastatic potential. Together, these results highlight the role of density-dependent synchrony dynamics as a fundamental, quantifiable mode of collective behavior in active epithelial matter, with implications for morphogenesis, cancer progression, and tissue diagnostics.

\section*{Introduction}
Collective cell migration plays a fundamental role in shaping the structural and functional properties of epithelial tissues, with implications spanning physiological processes—such as embryonic development and wound healing—to pathological conditions like cancer metastasis. Beyond its biological relevance, collective migration has emerged as a rich subject in non-equilibrium physics, inspiring new frameworks for understanding pattern formation, glassy dynamics, and active turbulence in living systems. In-vitro studies of mammalian epithelial monolayers have revealed a variety of striking collective phenomena, including flocking transitions~\cite{shen2025flocking}, active wetting \cite{perez2019active}, shear-induced responses in endothelial layers~\cite{destefano2017real}, mechanochemical pattern formation \cite{boocock2023interplay}, division-driven vortical flows \cite{rossen2014long}, collective mechanical waves \cite{tambe2011collective, serra2012mechanical}, and spontaneous swirling or rotational modes in confined domains~\cite{peyret2019sustained,ascione2022collective}. These dynamics typically arise from cell-cell interactions and their modulation by environmental cues. 

In confluent epithelia—systems of particular relevance to morphogenesis, wound repair, and cancer invasion—cell migration is strongly influenced by local contractility and density variations. These factors lead to large-scale spatial correlations and heterogeneities, reminiscent of the dynamic glass-like behavior observed in disordered materials~\cite{angelini2011glass,bi2016motility,garcia2015physics}. For instance, at intermediate densities, fast-moving cells tend to migrate in groups whose size increases with density, analogous to dynamic heterogeneity in supercooled liquids near the glass transition~\cite{bengtzelius1984dynamics,sillescu1999heterogeneity,jung2024dynamic,angelini2011glass}. At higher densities, both velocity correlation length and dynamic heterogeneity exhibit a non-monotonic dependence on density~\cite{garcia2015physics}, reflecting a complex interplay between cell-cell adhesion (e.g., via E-cadherin) and cell-substrate interactions (e.g., via vinculin). These observations raise important questions about the emergence and loss of temporal synchronization in such systems.

In this study, we investigate the spatio-temporal dynamics of large-scale temporal oscillations in collective cell migration using two model systems: (i) Madin-Darby Canine Kidney (MDCK) epithelial monolayers spanning a wide post-confluent density range, and (ii) human breast cancer cell lines with varying degrees of invasiveness. While previous studies have focused on collective modes such as shear, vorticity, and translation—each of which conserve area—we concentrate here on dilatational modes, characterized by the divergence of the velocity field and corresponding to local expansions and contractions. These modes are not only dominant in the bulk of confluent monolayers, but also closely linked to morphogenetic processes like gastrulation~\cite{supatto2005vivo, bhattacharya2021strain} and somitogenesis \cite{miao2023reconstruction}, and importantly, are straightforward to quantify.

Our results reveal the dynamics of synchrony in the dilatational field at multicellular scales, bearing resemblance to  excitable systems on multiple scales, from reaction-diffusion waves in starfish eggs~\cite{tan2020topological}, to bioelectrical waves in the heart and brain~\cite{norman2006beyond,gray1998spatial}. The persistence and defect statistics of these oscillatory phase patterns vary non-monotonically with cell density, delineating a transition from pre-jammed to jammed states. Through theoretical modeling, we demonstrate that intrinsic temporal synchronization in epithelial layers correlates with the spatial scale of cell-cell interactions. Extending our analysis to breast cancer cell lines, we find that malignant cells exhibit more temporally persistent and spatially coherent phase patterns than their non-malignant counterparts. Together, these findings reveal a previously underappreciated, self-organized temporal dimension to epithelial collective behavior. They provide a new framework for interpreting the dynamics of cell layers in both physiological and pathological states, and suggest that phase-based analysis of dilatational modes may offer a powerful lens for probing the mechanical and regulatory principles of collective cell migration.

\section*{Results}
\subsection*{Spatiotemporal dilatation patterns show synchronous oscillations in normative epithelia}
\begin{figure}[H]
  \centering
  \includegraphics[scale=0.7]{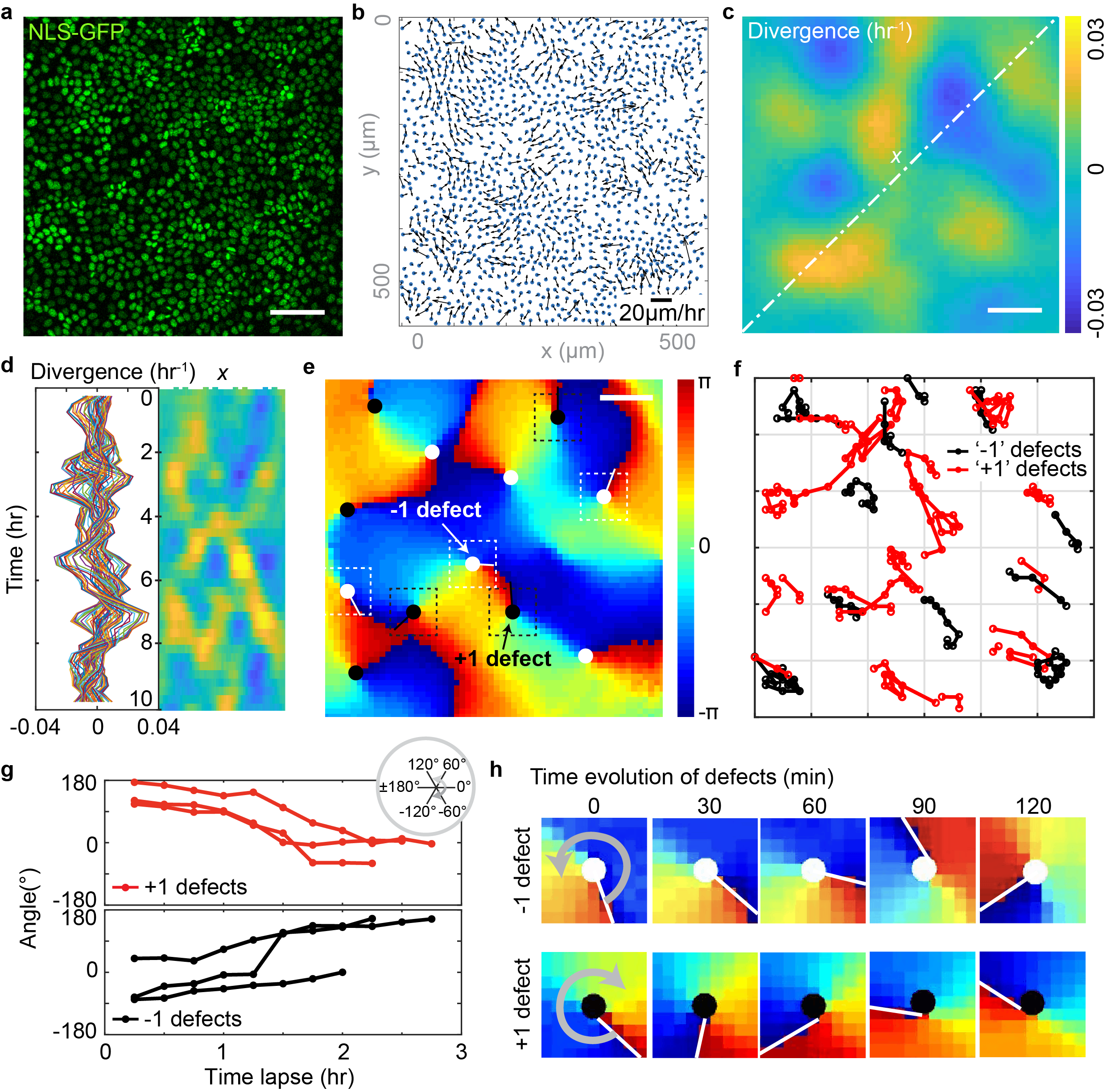}
  \caption{\textbf{Emergence of large-scale oscillatory patterns in MDCK cells on a flat surface.} \textbf{a}, NLS-GFP labeled cell nuclei. \textbf{b}, A snapshot of cell velocity calculated from nuclei displacement in consecutive frames. The length of velocity vectors is scaled with the velocity magnitude; Scale bar: 20 \(\mu\)m/hr. \textbf{c}, Velocity divergence $\boldsymbol{\nabla} \cdot \textbf{v}$ colormap. \textbf{d}, Left: velocity divergence variation in time for all the pixels along the dashed line in \textbf{c}. Right: kymograph of the velocity divergence show oscillatory ‘source’ (positive divergence) and ‘sink’ (negative divergence) patterns in time. Colorbar is the same as \textbf{c}. \textbf{e}, Divergence phase \(\phi(\textbf{x},t)=tan^{-1}(Div^*(\textbf{x},t),Div^*(\textbf{x},t+\tau))\) extracted from temporal divergence profile shows large-scale patterns, where \(\tau\) is one quarter of the period of divergence. Phase singularities are defined as defects. Black dots: +1 defects; white dots: -1 defects. \textbf{f}, Trajectories of +1 and -1 defects over 10hr. \textbf{g}, The angle change of 3 pairs of +1 and -1 defects (marked in the dashed boxed in \textbf{e}) over time. The angle is defined by the boundary of blue(\(-\pi\)) and red(\(\pi\)) phases. \textbf{h}, '-1' defects rotate anticlockwise and '+1' defects rotate clockwise. Scale bars in \textbf{a, c, e}: 50\(\mu m\).}
  \label{fig:1}
\end{figure}

MDCK cells are transfected with NLS-GFP (Green Fluorescent Protein tagged Nuclear Localization Signal) for visualization of cell nuclei and cultured on a 2D flat surface. The nuclei positions are recorded using a confocal microscope (Fig. \ref{fig:1}a), which allows us to track nuclear positions over time and calculate the velocity field \textbf{v}(\textbf{x},t) (Fig. \ref{fig:1}b). In addition to the visible flocking of cells which form packs, we also observe cell nuclei fluctuating back and forth when cell packs move and collide, creating oscillations within the cell monolayer which is consistent with findings from a previous study \cite{peyret2019sustained}. Indeed, an oscillatory expansion and contraction pattern in cell motion is visible from the raw video (Supplementary video 1). To quantify this behavior, we examine the velocity gradient which is itself decomposable into shear, vorticity and dilatational fields. On examining the shear and vorticity and their variation over the entire field of view, we find that the coefficient of variations (COV) of shear is significantly is smaller compared to the divergence (Fig. \ref{fig:S7}). This observation, together with the fact that only the dilatation mode causes local expansion and shrinking dynamics,  we focus on this mode which is characterized by spatio-temporal variations in the divergence (or dilatational) field \(Div \equiv \nabla\cdot \textbf{v}\).

In Fig. \ref{fig:1}c, we see that the dilatational field exhibits a periodic spatial pattern of sources (local positive peaks) and sinks (local negative peaks) with patterns spanning several tens of cells (Fig. \ref{fig:1}c). When cell clusters expand or contract, we observe that the local cell height decreases or increases respectively, as shown in Fig. \ref{fig:S6}. Interestingly, when we examine the temporal behavior of these peaks by plotting a kymograph of the divergence along the diagonal view, the sources and sinks also exhibit temporal periodicity (Fig. \ref{fig:1}d). The spatiotemporal oscillation in the divergence \(Div\) implies a dynamic interplay of cellular movement and pattern formation in cell density and divergence fields. 

To further investigate the temporal behavior of cell expansion and contraction, we calculate the local phase of the oscillation pattern following a previously proposed method \cite{gray1998spatial,tan2020topological} (see Methods in section \ref{methodsection}). For each pixel, we calculate the phase defined by \(\phi(\textbf{x},t)=tan^{-1}(Div^*(\textbf{x},t),Div^*(\textbf{x},t+\tau))\), where \(\tau\) is one quarter of the period of divergence signal over time and \(Div^*\) is the Gaussian-filtered detrended divergence \cite{gray1998spatial,tan2020topological}(see Method). As \(\tau\) does not vary significantly with position (Fig. \ref{fig:S1}c-d), we use the spatially averaged period \(\langle\tau\rangle_\textbf{x}\) to calculate the phase \(\phi(\textbf{x},t)\) for all pixels, in order to reduce the noise (see Methods section\ref{methodsection}). Remarkably, a large-scale divergence phase pattern emerges within the field of view (Fig. \ref{fig:1}e), indicating that the entire epithelial collective undergoes highly synchronized oscillations at a similar frequency. We further verify the phase calculation using Hilbert phase, which shows a consistent trend (Fig. \ref{fig:S2}). We notice the phase pattern \(\phi(\textbf{x},t)\) harbors a series of topological defects that generally have winding numbers of \(-1\) or +1, indicating local dynamic phase field rotating either clockwise or counterclockwise (Fig. \ref{fig:1}e,g,h). When we overlay the phase defects onto the velocity divergence colormap, they consistently localize at  regions of zero divergence (Fig. \ref{fig:S8}). Additionally, defects always appear and annihilate as pairs (Fig. \ref{fig:1}f) \cite{gray1998spatial}, due to topological constraints associated with the conservation of topological charge in two-dimensional. 

\subsection*{Cell density variations correlate with changes in synchrony}
 We next investigate the temporal patterns in the divergence field in MDCK cells with increasing cell number density (from \(1000 \:mm^{-2}\) to \(3500 \:mm^{-2}\)) (Fig. \ref{fig:2}a, d and Fig. \ref{fig:S3}b). As expected, the divergence magnitude decreases with density (Fig. \ref{fig:S3}a), due to the decreasing cell migration speed (Fig. \ref{fig:2}d) , consistent with Ref. \cite{angelini2011glass}.
Interestingly, the period of divergence, shown as its quarter \(\tau\) in both the colormap (Fig. \ref{fig:2}b) and the probability distribution (Fig. \ref{fig:2}e) , initially increases with density; however, it decreases again at large densities. Consistently, when we further quantify the phase of divergence \(\phi(\textbf{x},t)\) and plot its kymograph, it shows an increase in persistency as the density increases initially; and a decreases in the phase persistency at higher densities (Fig. \ref{fig:2}c). This is confirmed by calculating the temporal autocorrelation of the phase \(\phi\) and phase persistent time \(t_0\) as shown in Fig. \ref{fig:2}f, where \(t_0\) is determined from the time at which the autocorrelation crosses zero (see Methods). This non-monotonous trend in temporal behavior during cell density increase is reminiscent of the non-monotonic spatial velocity cross-correlation length reported previously in developing epithelia \cite{garcia2015physics}. Indeed, when we quantify the velocity cross-correlation length \(\zeta_{rr}\) (see Methods), we find that \(\zeta_{rr}\) first increases and then decreases as a function of cell density (Fig.~\ref{fig:2}g).

Similarly, the phase of the divergence  field exhibits discontinuities that give rise to positive and negative phase defects. We subsequently investigate how phase defects vary with cell density. Interestingly, the number of phase defects also shows a non-monotonic trend as the MDCK epithelia develop: In contrast to the trend observed in the persistence time, the number of defects first decreases and then increases (Fig. \ref{fig:2}h). These results demonstrate that the temporal collective cell migration, as evidenced by patterns in the divergence phase fields, is strongly influenced by changes in cellular density within the epithelial layer. The non-monotonic behavior of the velocity correlation (Fig. \ref{fig:2}g) suggests that cells transition from almost individual migration to collective migration with increasing density prior to a critical cell density, exhibiting more robust collective contractile and expansive oscillatory patterns. Beyond the critical density, cells are largely caged, and single-cell diffusion becomes a more prominent feature in cell motion, resulting in a decrease in temporal synchronization \cite{angelini2011glass}. 

Remarkably, this critical cell density has been demonstrated as a glass-transition density \(\rho_g\) when the collective cell migration speed converges with self-diffusion \cite{angelini2011glass}. In our system, we further observe that both the temporal persistence trend and the spatial velocity correlation length exhibit non-monotonic behavior as cell density increases, with turnover points occurring at similar densities (\(\sim1700\text{–}1800~\mathrm{mm}^{-2}\); Fig.~\ref{fig:2}f–g). As density increases, both the velocity correlation length and the phase persistence time first increase for densities below the transition density, and then decrease for densities above the transition density.
 As the cellular layer slows down when collective cell migration speed converges with self-diffusion, it behaves like an amorphous, glassy solid \cite{angelini2011glass}. This consistency in spatial and temporal trend indicates that there might be an intrinsic correlation between them.

\subsection*{A theory coupling active oscillations to cell density explains observations }
Inspired by the expansion and contraction dynamics observed in the experiments that show the spatio-temporal coupling of the cellular density to the active oscillations of the cells, we consider a minimal theory that couples these fields. As shown in the experimental analysis of the velocity gradient \(\boldsymbol{\nabla} \mathbf{v}\), the dominant contribution is the dilatational field (see SI Fig. S1). Therefore, we assume that the mechanical stress tensor associated with the movement of the confluent epithelial layer \(\boldsymbol{\sigma}\) arises from dilatation and expansion, so that we can  take \(\boldsymbol{\sigma} \sim \mathbf{I}\), neglecting shear contributions {(A comparison between different components of the velocity gradient tensor is given in Fig. \ref{fig:S7} in the SI.)}

To understand the dynamics of the cell monolayer as the density increases, we use a continuum description that couples the cell density $\rho$ with the the active contractile oscillations that may be characterized by a phase $\theta$, and an amplitude $A$ represented in terms of a complex field $G=A e^{i \theta}$.  
This suggests coupling the dilatation field expressed in terms of the divergence of the isotropic stress tensor associated with cellular expansion and contraction with an active oscillatory field, a theoretical framework that has been developed recently \cite{banerjee2024hydrodynamics}. We extend the model in~\cite{banerjee2024hydrodynamics} to allow for spatial variations in the phase and amplitude and thence consider the emergence and dynamics of topological defects in the phase field \(\theta\), leading to {a coupled Complex-Ginzburg-Landau {(CGL)} like theory~\cite{aranson2002world}. 
Then,  dynamics of the cell density $\rho({\bf x},t)$,  oscillation phase $\theta({\bf x},t)$, and amplitude $A{\bf x},t)$  are governed by the equations (details can be found in Supplementary Information, section \ref{sisection}) 
\begin{align}
&\partial_t \rho =  \alpha \nabla^2( \rho -\rho_0-\epsilon A \cos \theta),\label{eq1}\\
&\partial_t \theta = \Gamma (\nabla^2 \theta +2 \frac{\nabla A \cdot \nabla \theta}{A}) -\gamma  A \epsilon \sin \theta (\rho -\rho_0-A \epsilon \cos \theta) +\gamma \Omega\label{eq2},\\
& \partial_t A= \Gamma (\nabla^2 A -A (\nabla \theta)^2 )+ 2\chi A (\rho/\rho_0-A^2).\label{eq3}
\end{align}
Here $\rho_0$ represents the rescaled average density, and equation (1) describes how the diffusive dynamics of the local density field with diffusion coefficient $\alpha$, is driven by variations in the active contractility, while the (2-3) describe the coupling of the density of the cells to the usual {CGL} theory for the dynamics of spatio-temporal oscillations near a Hopf bifurcation (with a natural frequency $\Omega$ and diffusivity $\Gamma$). Furthermore, $\epsilon$ characterizes the coupling of the oscillatory field on the density, and $\chi$ characterizes the coupling of the density on the amplitude of the oscillatory field. When $\epsilon=\chi =0$,  we recover the usual {CGL} theory, but when these terms are non-vanishing, we get a system that is not rotationally invariant, as expected for situations where anisotropic effects are important (see SI for more details). We also note the presence of a natural length scale associated with the persistence of oscillations that follows by balancing the coefficient of the Laplacian $\Gamma$ with the homogeneous terms.}

To  understand the correlation between the spatial pattern and the temporal dynamics, we first assume no spatial dependency for the fields ($\rho=\rho_0$, and $\theta=\theta(t)$), and as such $\partial_t \rho=0$. Under this assumption, the dynamics of the phase reads
$\partial_t \theta = \frac{\epsilon^2}{2} \sin 2\theta  +\Omega$ which has an analytical solution given by $\theta=\cot ^{-1}\left(\frac{2 \Omega}{d \tan \left(D+d/2 \: t \right)-\epsilon^2}\right),$ where $D=\tan ^{-1}\left(\frac{\epsilon^2+2 \Omega \tan (\theta_0)}{d}\right)$, $d=\sqrt{4 \Omega^2-\epsilon^2}$, and $\theta_0$ is the initial phase. At long  times $t\gg D/d$, the initial condition becomes irrelevant and the period is given by $T=4 \pi/\sqrt{4\Omega^2-\epsilon^4}$. As a result, the presence of the nonlinear term $\epsilon^2$ in the dynamics of the phase increases the period of the oscilation from their intrinsic period $2 \pi/\Omega$. In the presence of spatial effects, assuming a constant amplitude, the coupled dynamics of the density and the phase read 
\begin{align}
&\partial_t \rho =  \alpha \nabla^2( \rho -\rho_0-\epsilon \cos \theta),\label{eq6}\\
&\partial_t \theta = -\epsilon \sin \theta ( \rho -\rho_0-\epsilon \cos \theta) +\Omega.\label{eq7}
\end{align}
The dynamics of the density in Eq. \eqref{eq6} suggests that density evolves to match the phase so that $\rho =\rho_0+\epsilon \cos \theta$. As a result the presence of the pattern in the density and phase decreases the effect of the first term in Eq. \eqref{eq7}, and as such decreases the period of the oscilations.

Going further, to understand the experimentally observed variations in the spatio-temporal persistence of the cells as a function of cellular densities,  we fit the experimental data for the spatial correlation as a function of density $\rho_0$, and use it as input to the model coefficient $\Gamma(\rho_0)$, which sets the spatial phase pattern (see SM, Section~\ref{fitsec}). Equations~\eqref{eq1}--\eqref{eq3} are then numerically solved for different values of cell density $\rho_0$. The fit to the experimental data results in an increased extent of the spatial patterns at the transition density $\rho_0 = 1$ (Fig.~\ref{figsim1}a,c and Fig. \ref{fig:S9}). Interestingly, the average oscillation period and average defect density as functions of cell density exhibit trends similar to those observed in experiments (Fig.~\ref{figsim1}d-e). Furthermore, in agreement with the experiments, $+1$ ($-1$) phase defects rotate clockwise (anticlockwise) over time (Fig.~\ref{figsim1}b). These measurements suggest that the temporal and spatial patterns are correlated, as expected from the simple analysis above.

\subsection*{Malignant cells have more persistent oscillation in breast cancer cell lines}
\begin{figure}[H]
\centering
  \includegraphics[scale=0.68]{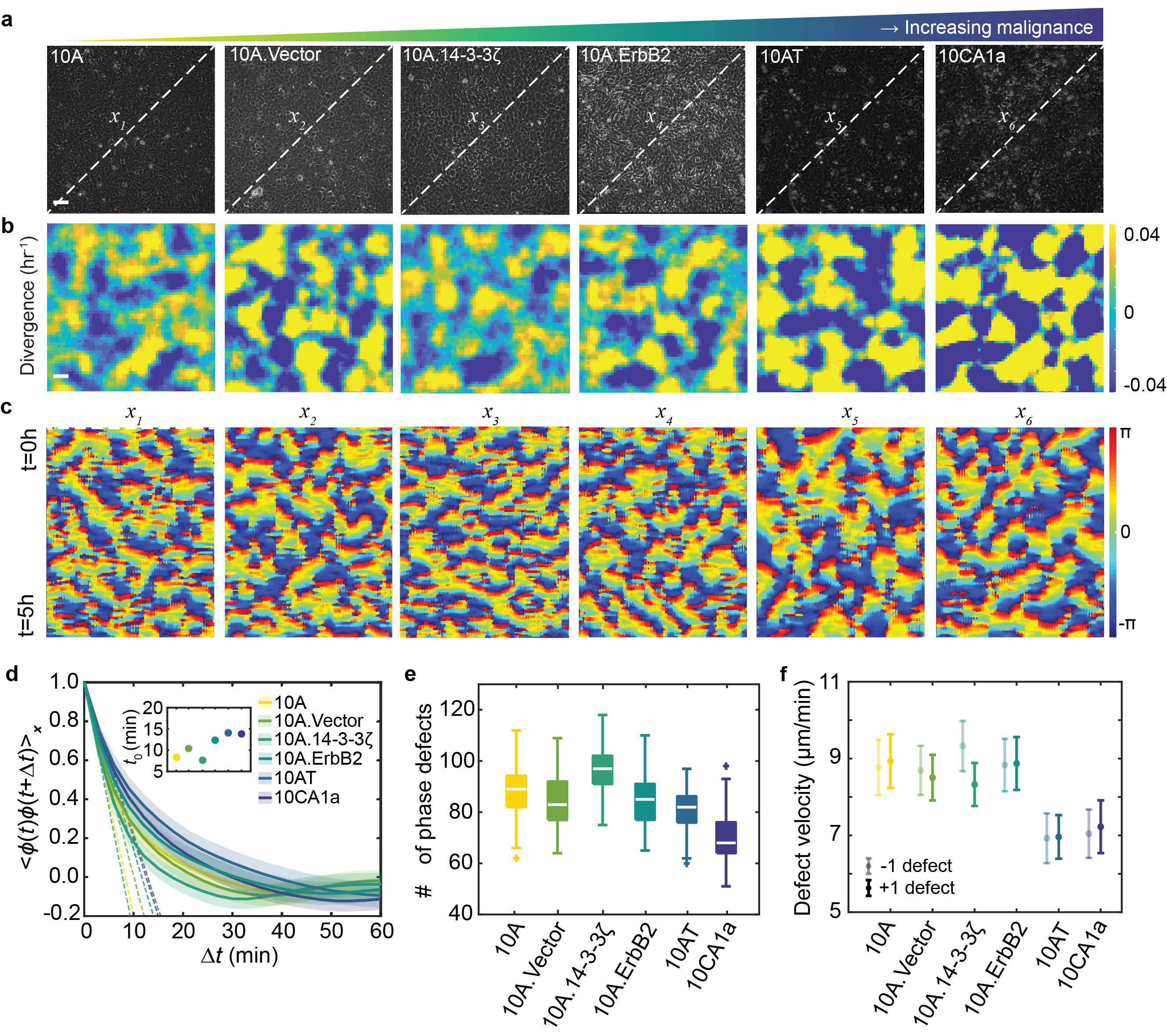}
  \caption{\textbf{The phase dynamics of divergence in breast cancer cell lines with increasing invasive potential at constant density.} \textbf{a}, Phase contrast images of breast cancer cell lines at constant density $(1900-2000) mm^{-2}$ with increasing invasive potential: 10A, 10.Vector, 10a.14-3-3\(\zeta\), 10A.ErbB2, 10AT, 10CA1a. \textbf{b}, The corresponding velocity divergence. In \textbf{a} and \textbf{b} scale bar is 100\(\mu m\). \textbf{c}, Phase \(\phi\) kymograph for 10A, 10.Vector, 10a.14-3-3\(\zeta\), 10A.ErbB2, 10AT, 10CA1a over 5hr, along the lines $x_1-x_6$. \textbf{d}, Time autocorrelation of  the phase \(\phi\). Inset: Persistence time \(t_0\) calculated as the time lag when autocorrelation function reaches zero. \textbf{e}, Box plot showing the number of defects in the phase decreases as cell lines become more malignant. \textbf{f}, Defect velocity decreases in the sequence of 10A, 10.Vector, 10a.14-3-3\(\zeta\), 10A.ErbB2, 10AT, 10CA1a cells. Error bars represent standard deviation over time.}
  \label{fig:3}
\end{figure}

To explore whether this dynamical analysis is informative to pathological processes such as cancer metastasis, we apply this method to a panel of \textit{in vitro} cultures of human breast cancer model systems, in the order of roughly increasing invasion potential as previously reported: MCF10A, MCF10A.Vector, MCF10A.14-3-3\(\zeta\), MCF10A.ErbB2, MCF10AT, MCF10CA1a \cite{kim2020unjamming}. For simplicity, in the rest of this manuscript we will denote these model systems as: 10A, 10A.Vector, 10A.14-3-3\(\zeta\), 10A.ErbB2, 10AT, 10CA1a, respectively, as shown in Fig. \ref{fig:3}a. 10AT and 10CA1a cell lines were derived by forced expression of mutated H-Ras followed by repeated selection from xenograft tumors. MCF10A.Vector, MCF10A.ErbB2, and
MCF10A.14–3-3\(\zeta\) cell lines were previously generated by stable transfection of control vector,
ErbB2, and 14–3-3\(\zeta\); over expression of ErbB2 or 14–3-3\(\zeta\) has been associated with
metastatic recurrence in breast cancer patients \cite{lu200914}. Unexpectedly, across different cell lines higher speeds are associated with enhanced size of cooperative cell packs \cite{kim2020unjamming}. 

For the six breast cancer cell lines, velocity field was previously obtained using particle image velocimetry (PIV) of cell phase contrast time-lapse images at similar densities $~(1900-2000)mm^{-2}$. Here, we calculate velocity divergence \(\nabla\cdot \textbf{v}\) using the velocity PIV field. Previous work has reported that the MCF10CA1a cell line exhibits significantly faster migratory speed relative to MCF10AT and MCF10A \cite{kim2020unjamming}. The velocity divergence shows that the most malignant cell lines, 10AT and 10CA1a, indeed have the largest magnitude of divergence, while the other four cell lines show much smaller velocity divergence \cite{kim2020unjamming} (Figs. \ref{fig:3}b and \ref{fig:S5}a-c). 
In addition, the average divergence within the field of view is around zero for the relatively benign cells (10A, 10A.Vector, 10A.14-3-3\(\zeta\), and 10A.ErbB2) but becomes positive for more invasive cells 10AT and 10A.CA1a (Fig. \ref{fig:S5}a). This might indicate that more malignant cell lines proliferate at a higher rate. Interestingly, the divergence coefficient of variation (COV)—defined as the standard deviation divided by the mean of the divergence signal—decreases with increasing malignancy, suggesting more homogeneous spatial divergence distribution (Fig. \ref{fig:S5}b). 

Using the same method introduced earlier, we then calculate the phase of divergence for these breast cancer cells (Fig. \ref{fig:S4}).  The phase \(\phi\) kymographs for different cell lines suggests that the persistent time of the phase increases as malignancy increases (Fig. \ref{fig:3}c). To quantify this, we perform a temporal autocorrelation of phase field and find that the phase persistent time increases with malignancy (Fig. \ref{fig:3}d, inset) in agreement with the phase kymograph. 
The increase in persistence time with malignancy is consistent with the previously reported trend of persistence length and cooperative cell pack size \cite{kim2020unjamming}, showing that more malignant cancer cells are potentially more persistent and collective to facilitate cancer invasion. To further explore temporal behaviors for cancer cells, we measure the number of phase defects. The measurements reveal that the density of the phase defects decreases with malignancy (Fig. \ref{fig:3}e). In addition, the phase defects have lower speed in more invasive cells (Fig. \ref{fig:3}f). Together, our temporal analysis reveals that more cancerous cells oscillate more persistently and harbor less phase defects. 

\newpage
\section*{Discussion}
While spatial correlations in collective cell motion have become a powerful tool for probing the non-equilibrium physics of epithelial tissues, our findings highlight the equally critical role of temporal correlations. We show that confluent epithelial monolayers exhibit large-scale, synchronized oscillations in the divergence of the velocity field—signatures of sustained expansion–contraction cycles at multicellular scales. These temporally coordinated dynamics emerge intrinsically from the collective behavior of cells, modulated by factors such as cell density, migration speed, and mechanical interactions.

A key result of our study is the identification of a non-monotonic dependence of oscillation period on cell density. Notably, both temporal persistence and spatial correlation length of divergence patterns peak at intermediate densities, then decline as density increases further. This mirrors trends reported for spatial correlations in previous studies, which link the onset of glassy behavior to a transition from directed to diffusive-like cell motion~\cite{angelini2011glass}. Below the glass transition density, cells engage in collective, coherent migration; above it, motion becomes increasingly constrained and randomized, resembling that of densely packed deformable particles. The parallel trends in spatial and temporal metrics observed here suggest that oscillatory phase dynamics offer an independent, and potentially more accessible, means to detect such glass-like transitions, particularly in systems where spatial resolution is limited.

To mechanistically interpret these observations, we formulated a continuum model inspired by the Complex Ginzburg–Landau framework. In this model, the cell density, and the phase, and amplitude of active oscillations are dynamically coupled. Local feedback between density and phase promotes phase alignment with a preferred density state, yielding spatial phase patterns while conserving global density. This local density alignment gives rise to spatial patterns in both density and phase (reducing the phase correlation length scale), and diminishes the density-driven feedback on the phase, resulting in faster oscillations with shorter periods, consistent with our experimental observations. Thus, our model captures how collective mechanical feedback can spontaneously generate and regulate temporally coherent behavior in epithelial sheets.

Extending these insights to pathological systems, we analyzed human breast cancer cell lines with increasing invasive potential ~\cite{kim2020unjamming,santner2001malignant,dawson1996mcf10at,imbalzano2009increasingly}. We find that highly malignant lines exhibited more persistent oscillations and fewer phase defects than benign or non-invasive counterparts. This temporal coherence may reflect an enhanced capacity for collective organization during invasion and metastasis, consistent with recent reports that collective migration persists even in the absence of strong cell-cell junctions~\cite{ilina2020cell,fischer2015epithelial}. Indeed, this persistency could facilitate more efficient tissue penetration and metastatic dissemination~\cite{friedl2012classifying}. Interestingly, re-analysis of previously reported datasets~\cite{kim2020unjamming} reveals a correlation between cell pack size and phase persistence: larger collectives show more temporally sustained dynamics. This reinforces a broader principle suggested by our findings—that temporal coherence in dilatational patterns tracks with spatial organization in both developmental and disease contexts.

Altogether, our study reveals that temporal phase dynamics in the dilatational modes of collective cell motion are not only robust and quantifiable, but also informative about the mechanical and organizational state of epithelial tissues. These patterns represent a previously underexplored, self-organized mode of collective behavior. By offering a dynamic, time-resolved complement to spatial correlation measures, temporal phase analysis holds promise for probing morphogenetic processes, characterizing disease states, and ultimately informing both diagnostic and therapeutic strategies in cancer biology.

\section*{Methods}\label{methodsection}
\subsection*{Experimental methods}
\textbf{MDCK cells imaging.} NLS-GFP MDCK cells are grown on PDMS flat surface under normal growth condition. Time resolution of MDCK cell migration video is 15min. MDCK cell velocity is calculated from fluorescent nuclei tracking use Trackmate.

\textbf{Breast cancer cell lines. } The experiments of breast cancer cell lines are from previous work by Kim et al \cite{kim2020unjamming}. MCF10A and other breast cancer cell lines’ velocity field is calculated using PIV from phase contrast images. The resolution of MCF10A and other breast cancer cell lines videos is 3min. MCF10A cell line and its derivatives, MCF10AT and MCF10CA1a cell lines, were derived by forced expression of mutated H-Ras followed by repeated selection from xenograft tumors; these cell lines exhibit increasingly transformed phenotypes and are well-characterized models of breast carcinoma progression \cite{santner2001malignant,dawson1996mcf10at,imbalzano2009increasingly}. In addition, MCF10A.Vector, MCF10A.14-3-3\(\zeta\), MCF10A.ErbB2 cell lines, which were generated by stable transfection of control vector, ErbB2; overexpression of ErbB2 or 14-3-3\(\zeta\) has been associated with metastatic recurrence in breast cancer patients \cite{lu200914}. Together, these breast carcinoma cell lines provide reliable \textit{in vitro} models with diverse levels of invasiveness \cite{santner2001malignant,dawson1996mcf10at,imbalzano2009increasingly,lu200914,baker2011cancer,pathak2013transforming,so2012differential}. 
Six breast carcinoma model cell lines exhibited diverse levels of cell-cell adhesion and EMT marker proteins. At day 3 of cell culture, E-cadherin protein was detected in all cell lines tested with the exception of 10A.14-3-3\(\zeta\) cells; the order of increasing E-cadherin expression was: 10A.14-3-3\(\zeta\)\(<\)10A.ErbB2 \(<\)10AT \(<\)10CA1a\(<\)10A\(<\)10A.Vector. Vimentin was detectable the strongest in 10A.14-3-3\(\zeta\) cells and was also detected in other carcinoma cell lines but not in 10A.Vector; the order of increasing vimentin expression was 10A.Vector \(<\) 10A \(<\) 10AT \(<\) 10CA1a \(<\) 10A.ErbB2 \(<\) 10A.14-3-3\(\zeta\). N-cadherin protein was detected the strongest in 10A.14-3-3\(\zeta\) cells and was also detected in other carcinoma cell lines.

\textbf{Divergence calculation.} Divergence is calculated based on previous work \cite{zehnder2015multicellular,tang2022collective}. We choose kernel radius to be R = 70\(\mu m\), the band width = 40\(\mu m\), grid size = 10\(\mu m\).

\textbf{Phase diagram calculation.} The spatiotemporal phase diagram \(\phi(\textbf{x},t)\) is calculated for each grid over time following the methods in Ref. \cite{gray1998spatial,tan2020topological}. We defined \(\phi(\textbf{x},t)=tan^{-1}(Div^*(\textbf{x},t),Div^*(\textbf{x},t+\tau))\), where \(Div^*(\textbf{x},t)=Div(\textbf{x},t)-\langle Div(\textbf{x},t)\rangle_t\) is the original divergence value subtract by the time averaged divergence and \(\tau\) is a quarter of the divergence signal’s period. To quantify \(\tau\), we first applied a Gaussian filter with the width of 5 time points to \(Div^*(\textbf{x},t)\) for each pixel (Fig. \ref{fig:S1}). Then we find time points when \(Div^*(\textbf{x},t)\) crosses zero. The half period is thus the average value of the time between every consecutive time points. Hilbert phase calculation was performed using the Hilbert function in MATLAB.

\textbf{Phase defects tracking.} To track phase defects migration, we used Trackmate in Fiji software, with linking max distance 50\(\mu m\).

\textbf{Phase Auto-correlation.} The autocorrelation of phase for each pixel is calculated using the autocorr function in MATLAB and averaged over space using $\langle \phi(\textbf{r},t)\phi(\textbf{r},t+\Delta t) \rangle_{\textbf{r}}$ where $\langle \rangle_\textbf{r}$ shows average in space. The persistent time of the phase \(\phi\) is estimated by linearly fitting the first 5 data points of the autocorrelation of the phase for the cancer cell data and finding the point where the line crosses zero for developing epithelia. 

\textbf{Defect instantaneous speed.} Instantaneous defects speed is calculated as \(\textbf{v}(t)=|\textbf{x}(t)-\textbf{x}(t-\Delta t)|/\Delta t\), where \(\Delta t = 3min\) is the is the time resolution of the video and $\textbf{x}(t)$ shows the position of the defect at time $t$. The average defect velocity is calculated by averaging the velocities of all defects over the duration of each experiments.

\textbf{Velocity cross-correlation length.} The velocity cross correlation is calculated using \(C_{vv}(r)=\langle \textbf{v}(\textbf{x},t)/|\textbf{v}(\textbf{x},t)|\cdot \textbf{v}(\textbf{x}+\Delta \textbf{x}, t)/|\textbf{v}(\textbf{x}+\Delta \textbf{x}, t)|\rangle_{\textbf{x},t,|\Delta \textbf{x}|=r}\). The velocity cross-correlation length is defined as the distance where the correlation function \(C_{vv}(r)\) reaches 0.1.

\subsection{Simulation Methods and Parameters}\label{secequations}
We solved the coupled equations \eqref{eq1}-\eqref{eq3} in a square with periodic boundary condition using a finite difference method with discretization length scales $\Delta x=\Delta y=1$, and discretization time scale $\Delta t=1$ on a $150 \times 150$, lattice over a total time of $t=1500000 \Delta t$. We used $\alpha=0.1$, $0.7\leq\epsilon\leq1$, $\Gamma_0=5\times 10^{-5}$, $\chi=10^{-4}$, $\gamma=10^{-3}$, $\Omega=1$, and we varied the density in the interval $0.7\leq\rho_0\leq2.2$. Our results stay valid for other choices of parameters, as long as the natural frequency $\Omega$ is large so that the patterns form in the system due to the mutal effect of $\Omega$ and $\epsilon$.  \\
The initial condition in all the simulations is a homogeneous density  $\rho=\rho_0$
with randomly chosen phase $\theta$, and constant amplitude $A=1$.
The initial phase is chosen from a random uniform distribution in the interval $[0, 2 \pi]$. The measurements are performed in steady state which is defined as a time when the number of defects fluctuates around a constant value for more than 1800 time steps. \\
The phase correlation length in Fig. \ref{figsim1}(c) is defined by measuring the spatial correlation of the phase vector $\textbf{P}(\textbf{r})=(\sin \theta(\textbf{r}),\cos \theta(\textbf{r}))$ using the definition of the spatial correlation function $c(r_0)= \langle \textbf{P}(\textbf{r})\cdot \textbf{P}(\textbf{r}+\textbf{r}_0)\rangle_{\textbf{r},t}$, where $\langle\rangle_{\textbf{r},t}$ shows the spatial and temporal average, and $r_0$ shows the distance between two grids. The correlation length is then defined as the length scale over which this function first crosses zero. \\
To find the average period in Fig. \ref{figsim1}(c),  we calculate the average rotational frequency defined as $\omega= \langle \Delta \theta(\textbf{r},t) /\Delta t\rangle_{\textbf{r},t}$, where $\Delta \theta(\textbf{r},t)$ shows the change in the phase between two consequnce time steps $t$ and $t+\Delta t$ at position $\textbf{r}$. We then define and compute the period as $T=2\pi/\omega$.

\section*{Author Contributions}
W.T. and M.G. conceptualized the project. W.T. performed experiments and experimental data analysis. M.R.N. and L.M. framed the theory. M.R.N. performed the simulations. A.F.P. provided raw experimental files for the six cancer cell lines, and constructive suggestions; A.F.P. edited an early version draft. W.T., M.R.N., L.M., and M.G. analyzed the data and wrote the manuscript.  L.M. and M.G. supervised the project. 

\section*{Acknowledgement}
We thank Jorn Dunkel, Jeffery Fredberg, Alasdair Hastewell, Ludwig Hoffmann, and Sifan Yin for helpful discussions.  W.T. and M.G. thank MathWorks Engineering Fellowship, National Institute Health grant number 1R01GM140108. L.M. thanks the Simons Foundation and the Henri Seydoux Fund for partial financial support. M.G. thanks support from Sloan Research Fellowship and Korea US Collaborative Research Fund (RS-2024-00468873).

\printbibliography

@article{angelini2011glass,
  title={Glass-like dynamics of collective cell migration},
  author={Angelini, Thomas E and Hannezo, Edouard and Trepat, Xavier and Marquez, Manuel and Fredberg, Jeffrey J and Weitz, David A},
  journal={Proceedings of the National Academy of Sciences},
  volume={108},
  number={12},
  pages={4714--4719},
  year={2011},
  publisher={National Academy of Sciences}
}

@article{nejad2024stress,
  title={Stress-shape misalignment in confluent cell layers},
  author={Nejad, Mehrana R and Ruske, Liam J and McCord, Molly and Zhang, Jun and Zhang, Guanming and Notbohm, Jacob and Yeomans, Julia M},
  journal={Nature Communications},
  volume={15},
  number={1},
  pages={3628},
  year={2024},
  publisher={Nature Publishing Group UK London}
}

@article{ascione2022collective,
  title={Collective rotational motion of freely expanding T84 epithelial cell colonies},
  author={Ascione, Flora and Caserta, Sergio and Esposito, Speranza and Villella, Valeria Rachela and Maiuri, Luigi and Nejad, Mehrana R and Doostmohammadi, Amin and Yeomans, Julia M and Guido, Stefano},
  journal={J. R. Soc. Interface},
  volume={20},
  number={199},
  pages={20220719},
  year={2023},
  publisher={The Royal Society}
}

@article{lu200914,
  title={14-3-3$\zeta$ cooperates with ErbB2 to promote ductal carcinoma in situ progression to invasive breast cancer by inducing epithelial-mesenchymal transition},
  author={Lu, Jing and Guo, Hua and Treekitkarnmongkol, Warapen and Li, Ping and Zhang, Jian and Shi, Bin and Ling, Chen and Zhou, Xiaoyan and Chen, Tongzhen and Chiao, Paul J and others},
  journal={Cancer cell},
  volume={16},
  number={3},
  pages={195--207},
  year={2009},
  publisher={Elsevier}
}

@article{banerjee2024hydrodynamics,
  title={Hydrodynamics of pulsating active liquids},
  author={Banerjee, Tirthankar and Desaleux, Thibault and Ranft, Jonas and Fodor, {\'E}tienne},
  journal={arXiv preprint arXiv:2407.19955},
  year={2024}
}

@article{bi2016motility,
  title={Motility-driven glass and jamming transitions in biological tissues},
  author={Bi, Dapeng and Yang, Xingbo and Marchetti, M Cristina and Manning, M Lisa},
  journal={Physical Review X},
  volume={6},
  number={2},
  pages={021011},
  year={2016},
  publisher={APS}
}

@article{peyret2019sustained,
  title={Sustained oscillations of epithelial cell sheets},
  author={Peyret, Gr{\'e}goire and Mueller, Romain and d’Alessandro, Joseph and Begnaud, Simon and Marcq, Philippe and M{\`e}ge, Ren{\'e}-Marc and Yeomans, Julia M and Doostmohammadi, Amin and Ladoux, Beno{\^\i}t},
  journal={Biophysical journal},
  volume={117},
  number={3},
  pages={464--478},
  year={2019},
  publisher={Elsevier}
}

@article{boocock2023interplay,
  title={Interplay between mechanochemical patterning and glassy dynamics in cellular monolayers},
  author={Boocock, Daniel and Hirashima, Tsuyoshi and Hannezo, Edouard},
  journal={PRX Life},
  volume={1},
  number={1},
  pages={013001},
  year={2023},
  publisher={APS}
}

@article{rossen2014long,
  title={Long-range ordered vorticity patterns in living tissue induced by cell division},
  author={Rossen, Ninna S and Tarp, Jens M and Mathiesen, Joachim and Jensen, Mogens H and Oddershede, Lene B},
  journal={Nature communications},
  volume={5},
  number={1},
  pages={5720},
  year={2014},
  publisher={Nature Publishing Group UK London}
}

@article{norman2006beyond,
  title={Beyond mind-reading: multi-voxel pattern analysis of fMRI data},
  author={Norman, Kenneth A and Polyn, Sean M and Detre, Greg J and Haxby, James V},
  journal={Trends in cognitive sciences},
  volume={10},
  number={9},
  pages={424--430},
  year={2006},
  publisher={Elsevier}
}

@article{gray1998spatial,
  title={Spatial and temporal organization during cardiac fibrillation},
  author={Gray, Richard A and Pertsov, Arkady M and Jalife, Jos{\'e}},
  journal={Nature},
  volume={392},
  number={6671},
  pages={75--78},
  year={1998},
  publisher={Nature Publishing Group UK London}
}

@article{tan2020topological,
  title={Topological turbulence in the membrane of a living cell},
  author={Tan, Tzer Han and Liu, Jinghui and Miller, Pearson W and Tekant, Melis and Dunkel, J{\"o}rn and Fakhri, Nikta},
  journal={Nature Physics},
  volume={16},
  number={6},
  pages={657--662},
  year={2020},
  publisher={Nature Publishing Group UK London}
}

@article{garcia2015physics,
  title={Physics of active jamming during collective cellular motion in a monolayer},
  author={Garcia, Simon and Hannezo, Edouard and Elgeti, Jens and Joanny, Jean-Fran{\c{c}}ois and Silberzan, Pascal and Gov, Nir S},
  journal={Proceedings of the National Academy of Sciences},
  volume={112},
  number={50},
  pages={15314--15319},
  year={2015},
  publisher={National Academy of Sciences}
}

@article{kim2020unjamming,
  title={Unjamming and collective migration in MCF10A breast cancer cell lines},
  author={Kim, Jae Hun and Pegoraro, Adrian F and Das, Amit and Koehler, Stephan A and Ujwary, Sylvia Ann and Lan, Bo and Mitchel, Jennifer A and Atia, Lior and He, Shijie and Wang, Karin and others},
  journal={Biochemical and biophysical research communications},
  volume={521},
  number={3},
  pages={706--715},
  year={2020},
  publisher={Elsevier}
}

@article{ilina2020cell,
  title={Cell--cell adhesion and 3D matrix confinement determine jamming transitions in breast cancer invasion},
  author={Ilina, Olga and Gritsenko, Pavlo G and Syga, Simon and Lippoldt, J{\"u}rgen and La Porta, Caterina AM and Chepizhko, Oleksandr and Grosser, Steffen and Vullings, Manon and Bakker, Gert-Jan and Starru{\ss}, J{\"o}rn and others},
  journal={Nature cell biology},
  volume={22},
  number={9},
  pages={1103--1115},
  year={2020},
  publisher={Nature Publishing Group UK London}
}

@article{tambe2011collective,
  title={Collective cell guidance by cooperative intercellular forces},
  author={Tambe, Dhananjay T and Corey Hardin, C and Angelini, Thomas E and Rajendran, Kavitha and Park, Chan Young and Serra-Picamal, Xavier and Zhou, Enhua H and Zaman, Muhammad H and Butler, James P and Weitz, David A and others},
  journal={Nature materials},
  volume={10},
  number={6},
  pages={469--475},
  year={2011},
  publisher={Nature Publishing Group UK London}
}

@article{serra2012mechanical,
  title={Mechanical waves during tissue expansion},
  author={Serra-Picamal, Xavier and Conte, Vito and Vincent, Romaric and Anon, Ester and Tambe, Dhananjay T and Bazellieres, Elsa and Butler, James P and Fredberg, Jeffrey J and Trepat, Xavier},
  journal={Nature Physics},
  volume={8},
  number={8},
  pages={628--634},
  year={2012},
  publisher={Nature Publishing Group UK London}
}

@article{santner2001malignant,
  title={Malignant MCF10CA1 cell lines derived from premalignant human breast epithelial MCF10AT cells},
  author={Santner, Steven J and Dawson, Peter J and Tait, Larry and Soule, Herbert D and Eliason, James and Mohamed, Anwar N and Wolman, Sandra R and Heppner, Gloria H and Miller, Fred R},
  journal={Breast cancer research and treatment},
  volume={65},
  pages={101--110},
  year={2001},
  publisher={Springer}
}

@article{sillescu1999heterogeneity,
  title={Heterogeneity at the glass transition: a review},
  author={Sillescu, Hans},
  journal={Journal of Non-Crystalline Solids},
  volume={243},
  number={2-3},
  pages={81--108},
  year={1999},
  publisher={Elsevier}
}

@article{jung2024dynamic,
  title={Dynamic heterogeneity at the experimental glass transition predicted by transferable machine learning},
  author={Jung, Gerhard and Biroli, Giulio and Berthier, Ludovic},
  journal={Physical Review B},
  volume={109},
  number={6},
  pages={064205},
  year={2024},
  publisher={APS}
}

@article{bengtzelius1984dynamics,
  title={Dynamics of supercooled liquids and the glass transition},
  author={Bengtzelius, Ulf and Gotze, W and Sjolander, A},
  journal={Journal of Physics C: solid state Physics},
  volume={17},
  number={33},
  pages={5915},
  year={1984},
  publisher={IOP Publishing}
}

@article{perez2019active,
  title={Active wetting of epithelial tissues},
  author={P{\'e}rez-Gonz{\'a}lez, Carlos and Alert, Ricard and Blanch-Mercader, Carles and G{\'o}mez-Gonz{\'a}lez, Manuel and Kolodziej, Tomasz and Bazellieres, Elsa and Casademunt, Jaume and Trepat, Xavier},
  journal={Nature physics},
  volume={15},
  number={1},
  pages={79--88},
  year={2019},
  publisher={Nature Publishing Group UK London}
}

@article{dawson1996mcf10at,
  title={MCF10AT: a model for the evolution of cancer from proliferative breast disease},
  author={Dawson, Peter J and Wolman, Sandra R and Tait, Lany and Heppner, Gloria H and Miller, Fred R},
  journal={The American journal of pathology},
  volume={148},
  number={1},
  pages={313},
  year={1996}
}

@article{aranson2002world,
  title={The world of the complex Ginzburg-Landau equation},
  author={Aranson, Igor S and Kramer, Lorenz},
  journal={Reviews of modern physics},
  volume={74},
  number={1},
  pages={99},
  year={2002},
  publisher={APS}
}

@article{destefano2017real,
  title={Real-time quantification of endothelial response to shear stress and vascular modulators},
  author={DeStefano, Jackson G and Williams, Ashley and Wnorowski, Alexa and Yimam, Nahom and Searson, Peter C and Wong, Andrew D},
  journal={Integrative Biology},
  volume={9},
  number={4},
  pages={362--374},
  year={2017},
  publisher={Oxford University Press}
}

@article{shen2025flocking,
  title={Flocking and giant fluctuations in epithelial active solids},
  author={Shen, Yuan and O’Byrne, J{\'e}r{\'e}my and Schoenit, Andreas and Maitra, Ananyo and M{\`e}ge, Ren{\'e}-Marc and Voituriez, Rapha{\"e}l and Ladoux, Benoit},
  journal={Proceedings of the National Academy of Sciences},
  volume={122},
  number={16},
  pages={e2421327122},
  year={2025},
  publisher={National Academy of Sciences}
}

@article{imbalzano2009increasingly,
  title={Increasingly transformed MCF-10A cells have a progressively tumor-like phenotype in three-dimensional basement membrane culture},
  author={Imbalzano, Karen M and Tatarkova, Iva and Imbalzano, Anthony N and Nickerson, Jeffrey A},
  journal={Cancer cell international},
  volume={9},
  pages={1--11},
  year={2009},
  publisher={Springer}
}

@article{miao2023reconstruction,
  title={Reconstruction and deconstruction of human somitogenesis in vitro},
  author={Miao, Yuchuan and Djeffal, Yannis and De Simone, Alessandro and Zhu, Kongju and Lee, Jong Gwan and Lu, Ziqi and Silberfeld, Andrew and Rao, Jyoti and Tarazona, Oscar A and Mongera, Alessandro and others},
  journal={Nature},
  volume={614},
  number={7948},
  pages={500--508},
  year={2023},
  publisher={Nature Publishing Group UK London}
}

@article{bhattacharya2021strain,
  title={Strain maps characterize the symmetry of convergence and extension patterns during zebrafish gastrulation},
  author={Bhattacharya, Dipanjan and Zhong, Jun and Tavakoli, Sahar and Kabla, Alexandre and Matsudaira, Paul},
  journal={Scientific reports},
  volume={11},
  number={1},
  pages={19357},
  year={2021},
  publisher={Nature Publishing Group UK London}
}

@article{supatto2005vivo,
  title={In vivo modulation of morphogenetic movements in Drosophila embryos with femtosecond laser pulses},
  author={Supatto, Willy and D{\'e}barre, Delphine and Moulia, Bruno and Brouz{\'e}s, Eric and Martin, Jean-Louis and Farge, Emmanuel and Beaurepaire, Emmanuel},
  journal={Proceedings of the National Academy of Sciences},
  volume={102},
  number={4},
  pages={1047--1052},
  year={2005},
  publisher={National Academy of Sciences}
}

@article{fischer2015epithelial,
  title={Epithelial-to-mesenchymal transition is not required for lung metastasis but contributes to chemoresistance},
  author={Fischer, Kari R and Durrans, Anna and Lee, Sharrell and Sheng, Jianting and Li, Fuhai and Wong, Stephen TC and Choi, Hyejin and El Rayes, Tina and Ryu, Seongho and Troeger, Juliane and others},
  journal={Nature},
  volume={527},
  number={7579},
  pages={472--476},
  year={2015},
  publisher={Nature Publishing Group UK London}
}

@article{friedl2012classifying,
  title={Classifying collective cancer cell invasion},
  author={Friedl, Peter and Locker, Joseph and Sahai, Erik and Segall, Jeffrey E},
  journal={Nature cell biology},
  volume={14},
  number={8},
  pages={777--783},
  year={2012},
  publisher={Nature Publishing Group UK London}
}

@article{baker2011cancer,
  title={Cancer cell migration: integrated roles of matrix mechanics and transforming potential},
  author={Baker, Erin L and Srivastava, Jaya and Yu, Dihua and Bonnecaze, Roger T and Zaman, Muhammad H},
  journal={PLoS One},
  volume={6},
  number={5},
  pages={e20355},
  year={2011},
  publisher={Public Library of Science San Francisco, USA}
}

@article{pathak2013transforming,
  title={Transforming potential and matrix stiffness co-regulate confinement sensitivity of tumor cell migration},
  author={Pathak, Amit and Kumar, Sanjay},
  journal={Integrative Biology},
  volume={5},
  number={8},
  pages={1067--1075},
  year={2013},
  publisher={Oxford University Press}
}

@article{so2012differential,
  title={Differential expression of key signaling proteins in MCF10 cell lines, a human breast cancer progression model},
  author={So, Jae Young and Lee, Hong Jin and Kramata, Pavel and Minden, Audrey and Suh, Nanjoo},
  journal={Molecular and cellular pharmacology},
  volume={4},
  number={1},
  pages={31},
  year={2012}
}

@article{zehnder2015multicellular,
  title={Multicellular density fluctuations in epithelial monolayers},
  author={Zehnder, Steven M and Wiatt, Marina K and Uruena, Juan M and Dunn, Alison C and Sawyer, W Gregory and Angelini, Thomas E},
  journal={Physical Review E},
  volume={92},
  number={3},
  pages={032729},
  year={2015},
  publisher={APS}
}

@article{tang2022collective,
  title={Collective curvature sensing and fluidity in three-dimensional multicellular systems},
  author={Tang, Wenhui and Das, Amit and Pegoraro, Adrian F and Han, Yu Long and Huang, Jessie and Roberts, David A and Yang, Haiqian and Fredberg, Jeffrey J and Kotton, Darrell N and Bi, Dapeng and others},
  journal={Nature Physics},
  volume={18},
  number={11},
  pages={1371--1378},
  year={2022},
  publisher={Nature Publishing Group UK London}
}

\clearpage
\section*{Figures}

\begin{figure}[H]
\centering
  \includegraphics[scale=0.68]{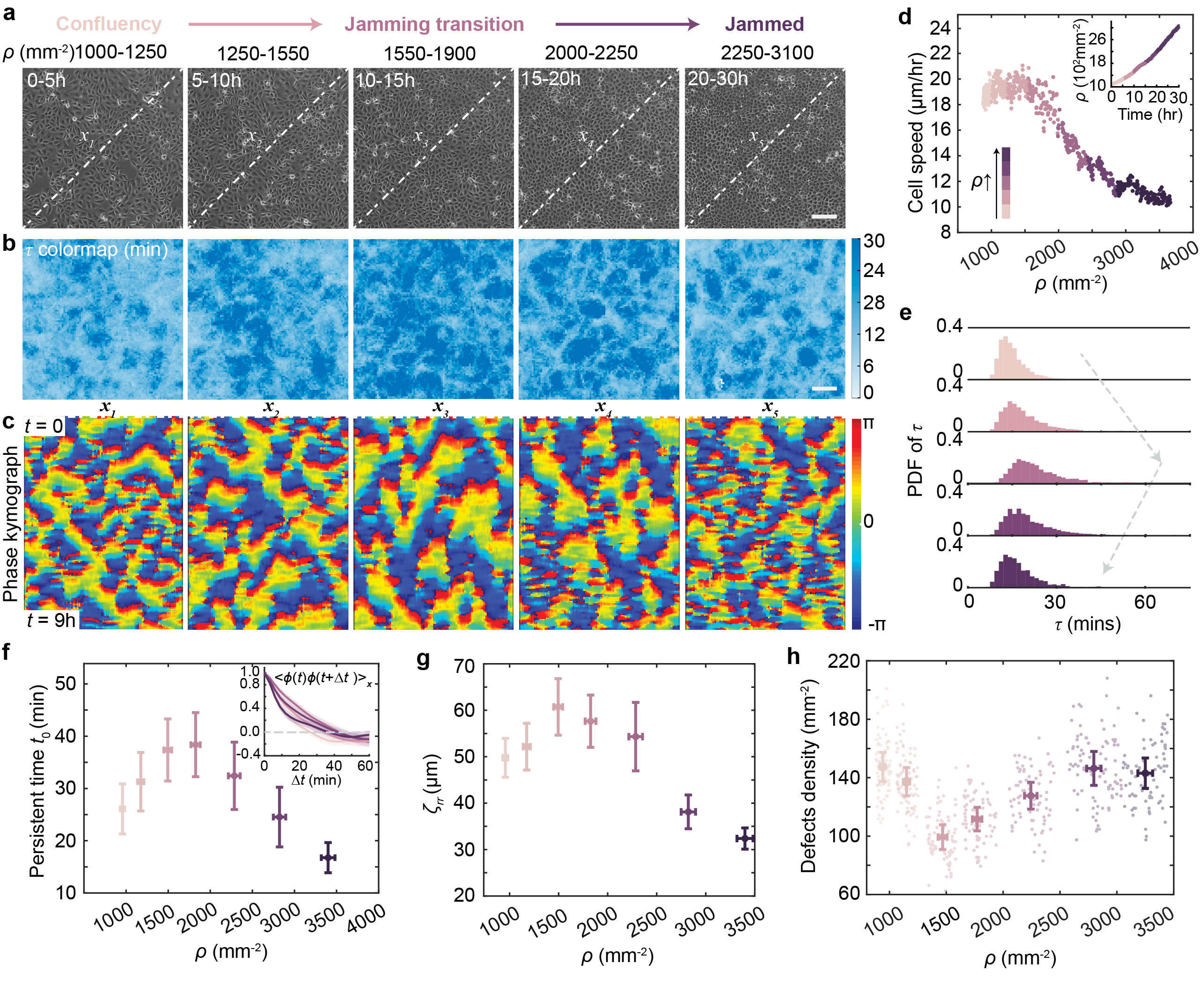}
  \caption{\textbf{Cell density affects temporal oscillations in developing MDCK cells.} \textbf{a,} Phase contrast images of the monolayer as density increases during 30 hr (from left to right). \textbf{b}, Spatial colormap of \(\tau\), where \(\tau\) is one quarter of divergence oscillation period. Scale bars in \textbf{a-b}: 50\(\mu m\). \textbf{c}, Kymographs of the phase \(\phi\) for divergence along \(x_1, x_2, x_3, x_4, x_5\) in \textbf{a}. Time interval between two frames is 3 minutes. \textbf{d}, Cell speed decreases as cell density increases. Inset: Cell density increases as a function of time. \textbf{e}, Probability density function (PDF) of the oscillation period \(\tau\). Bin size: 1.5 min. \textbf{f}, Temporal autocorrelation of \(\phi\) (inset) and the persistence time \(t_0\) at different densities. Dashed lines in the inset show the time lag when the autocorrelations reach zero to obtain the persistence time \(t_0\). \textbf{g}, Velocity cross-correlation length \(\zeta_{rr}\) as a function of density. \textbf{h}, Defects density in the phase as a function of density.}
  \label{fig:2}
\end{figure}

\begin{figure}[H]
	\centering
	\includegraphics[scale=0.41]{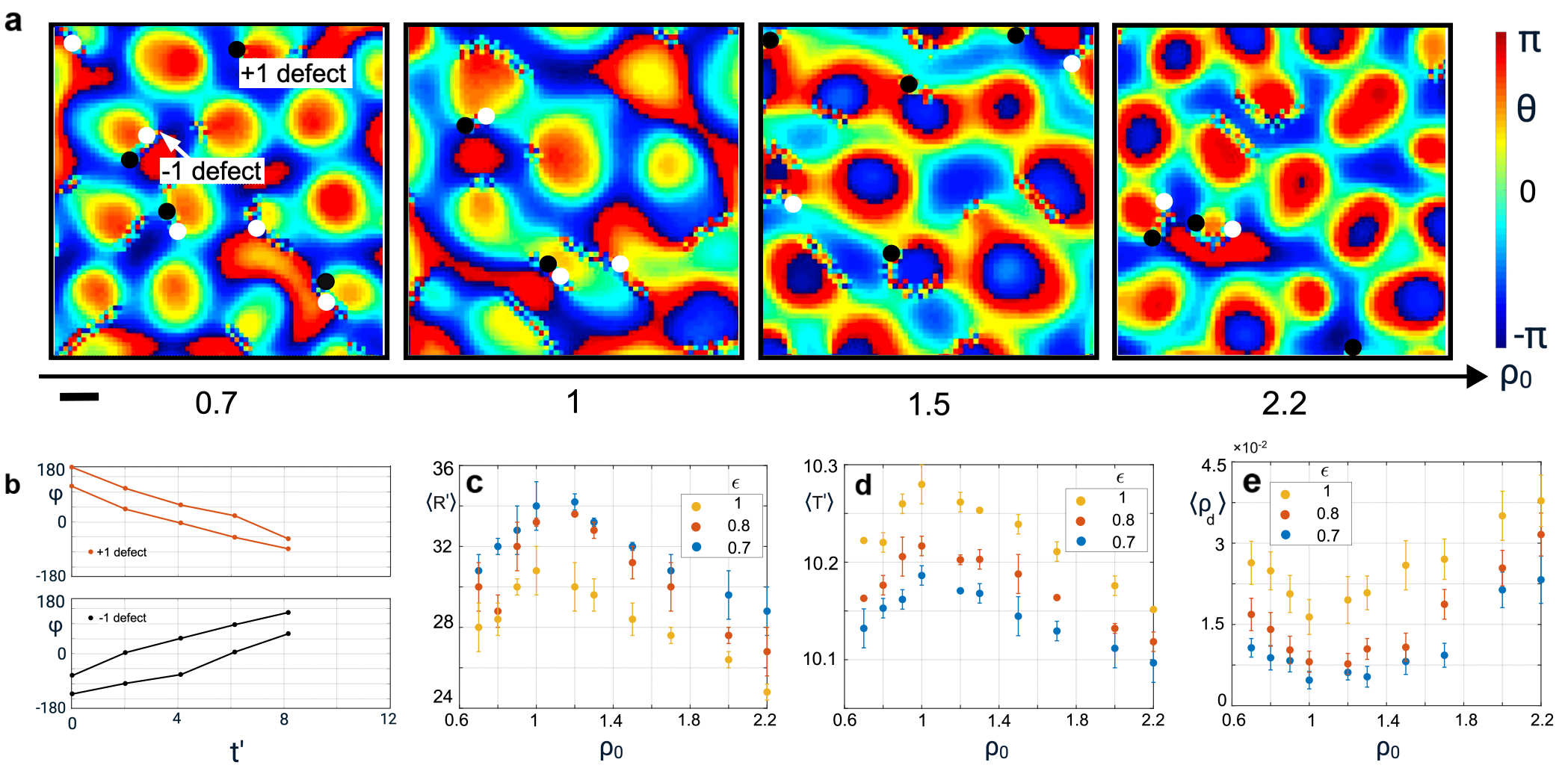}
	\caption{ \textbf{Continuum model recapitulates the non-monotonic temporal trend as a function of density when inputting the spatial interaction length scale.} \textbf{a}, Snapshots of simulations for various initial densities $\rho_0$. Increasing the initial density first increases and then decreases the pattern size in the phase. The scale bar represents 15 simulation grid units. $+1$ ($-1$) phase defects are shown in black (white). \textbf{b}, Top (bottom) shows two examples of rotation of $+1$ ($-1$) defect for $\rho_0=1$ in simulations as a function of dimensionless time $t'=t \Omega \gamma/(2 \pi)$. $\phi$ is defined as in experiments in Fig. \ref{fig:1}g.
\textbf{c}, Dimensionless phase correlation length $R' = (\gamma \Omega/ \Gamma_0)^{1/2} R$ as a function of density, where $R$ is the correlation length measured in simulation grid units. We fitted quadratic functions to the square of the velocity correlation length from experiments for densities below and above the transition density $\rho_0 = 1$ to determine the functional form of $\Gamma(\rho_0)$ as a function of $\rho_0$ and considered $\Gamma$ to be proportional to the fit with a proportionality constant $\Gamma_0$ (see Section~\ref{fitsec} in the SI). The simulation results shown here yield a spatial correlation consistent with experimental observations.
    \textbf{d},  
     Non-dimensionalised average period of the oscillations $\langle T'\rangle$ as a function of density, where $T'= T \Omega \gamma/(2 \pi)$, and $T$ is the period of the oscillation measured in simulation time step units. The average period shows a peak around  the density $\rho_0=1$ where the correlation length is maximum, in agreement with the experiments. \textbf{d}, Average defect density $\rho_d' = \Gamma_0 \rho_d/(\gamma \Omega)$, where $\rho_d$ is the average defect density in terms of simulation grid unit, as a function of density. Defect density first decreases and then increases with cell density.  Parameter values are defined in the SI. In panels (a–e), the mean and standard deviation for each density are calculated over four simulations.}
\label{figsim1}
\end{figure}

\newpage
\section{Supplementary Information}\label{sisection}
\renewcommand{\thepage}{S\arabic{page}} 
\renewcommand{\theequation}{S\arabic{equation}} 
\renewcommand{\thesection}{S\arabic{section}}  
\renewcommand{\thetable}{S\arabic{table}}  
\renewcommand{\thefigure}{S\arabic{figure}}
\setcounter{equation}{0}
\setcounter{page}{1}
\subsection{Components of velocity gradient tensor}
{The velocity gradient tensor $\boldsymbol{\nabla \mathbf{v}}$ can be decomposed into the rate of shear tensor $\boldsymbol{\Upsilon}$, the dilatation $Div = \boldsymbol{\nabla} \cdot \mathbf{v}$, and the vorticity $\boldsymbol{\Omega}$, defined in 2D as:
\begin{align}
    &\boldsymbol{\Upsilon}=(\boldsymbol{\nabla} \textbf{v}^T+\boldsymbol{\nabla} \textbf{v}-\textbf{I}\boldsymbol{\nabla}\cdot \textbf{v})/2,\\
    &Div=\boldsymbol{\nabla}\cdot \textbf{v},\\
    &\boldsymbol{\Omega}=(\boldsymbol{\nabla} \textbf{v}^T-\boldsymbol{\nabla} \textbf{v})/2,
\end{align}
where $\mathbf{I}$ denotes the identity tensor.  
We measured both the magnitude and the Coefficient of Variation (COV) of the different components of the velocity gradient in the experiments over time (see Fig.~\ref{fig:S7}). As expected, the magnitude of all components decreases with increasing cell density, due to a reduction in the overall velocity. Although all components are of a similar order of magnitude (Fig.~\ref{fig:S7}(a)), the COV is significantly larger for the velocity divergence and the vorticity (Fig.~\ref{fig:S7}(b)), suggesting that spatial patterns predominantly emerge in these fields.
}\\
{In this paper, we focus on cell density dynamics and modulation. Accordingly, we neglect both vorticity, which leads to cell rotation without affecting the local density, and shear, which results in cell deformation in the absence of changes in the density (or area). We therefore restrict our analysis to the dynamics and patterns associated with dilatational modes, leaving aside  the evolution of a tensor field associated with the orientation of force-generating agents and cell shape, as addressed for example in \cite{nejad2024stress}.
}
\subsection{Model}\label{simodel}
To construct a continuum theory that couples the cell density and the complex-valued order parameter describing the active oscillations of the cells, we follow~\cite{banerjee2024hydrodynamics} and  phenomenologically write the coupled dynamics of the density $\rho({\bf x},t)$ and the complex field $ {G}({\bf x},t)$ as
\begin{align}
&\partial_t \rho = \alpha \nabla^2 (a_1+a_2 (\bar{G}+G)+a_3 (\bar{G}-G)),\label{eq8}\\
&\partial_t G = b_1 \nabla^2 G+ b_2 G |G|^2+ b_3 G + b_4 G (G+\bar{G}) + b_5 G (G-\bar{G})+ b_6 G (G-\bar{G})(G+\bar{G})\nonumber\\&+b_7 (G-\bar{G}) (G+\bar{G})+b_8 \nabla \rho \cdot \nabla G+b_9 \nabla \rho \cdot \nabla \bar{G},\label{eq9}
\end{align}
where $a_i=a_i(\rho)$ and $b_i=b_i(\rho)$ are functions of the density. The absence of $\nabla^2 \bar{G}$ is a consequence of symmetry principles, and rotational invariance in the absence of activity (expressed as a dependence on the density). Similarly, we reject terms such as $\bar{G} G^2$ or $\bar{G}$ as they do not satisfy rotational invariance. Then, in the absence of activity, we recover the usual CGL. The coefficients $a_2,a_3,b_4$, $b_5$, $b_6$ and $b_9$ break rotational invariance via the coupling of the  amplitude and phase to the density.
We chose the coefficients so that our equations reduce to those introduced in Ref.~\cite{banerjee2024hydrodynamics} for a constant amplitude $A$. Accordingly, we set $a_3 = b_4 = b_7 = b_8 = b_9 = 0$, and thus we have:

  \begin{align}
&\partial_t \rho = \alpha \nabla^2 (a_1+2 a_2 A \cos \theta),\label{eq10}\\
&\partial_t G = \Gamma \nabla^2 G+ G (b_2 A^2+ b_3 ) + 2 i  G A [b_5 \sin \theta+  b_6 A  \sin 2\theta]. \label{eq11}
\end{align}
Equations~\eqref{eq10} and \eqref{eq11} can be expanded and written in terms of the dynamics of the phase and amplitude as
  \begin{align}
  &\partial_t \rho = \alpha \nabla^2 (a_1+2 a_2 A \cos \theta),\label{eq12}\\
&\partial_t \theta = 2A \Re(b_5 \sin\theta+b_6 A \sin2\theta)+\Im{(b_2 A^2 +b_3)}\label{eq13},\\
&\partial_t A =A\Re{(b_2 A^2 +b_3)}-2 A^2 \Im{(b_5 \sin\theta+b_6 A \sin2\theta)}.\label{eq14}
\end{align}
In the dynamics of the density, the term with coefficients $a_2$ breaks the gauge invariance.
When $A=1$, these equations converge to the equations introduced in Ref. \cite{banerjee2024hydrodynamics} for 
\begin{align}
&a_1=\rho -\rho_0,\:\:\:
a_2=-\epsilon/2, \:\:\: b_7=0, \:\: b_1=\Gamma, \:\:\: b_2=-\chi, \:\:\: b_3= (i (\gamma \Omega)+\rho \chi), \:\:\: b_5=\gamma \epsilon (1-\rho)/2, \:\:\: \nonumber \\ &b_4=b_7=0,\:\:\: b_6=\gamma \epsilon^2/4, \:\:\: \label{eq15}
\end{align}
where $\rho_0$ is the homogeneous density. Using the coefficients in Eq. \eqref{eq15}, the dynamics of the phase, amplitude and the density are governed by the minimal model introduced in Eqs. \eqref{eq1}-\eqref{eq3} in the manuscript
\begin{align}
&\partial_t \rho =  \alpha \nabla^2( \rho -\rho_0 -\epsilon A \cos \theta),\label{eq16}\\
&\partial_t \theta = \Gamma (\nabla^2 \theta +2 \frac{\nabla A \cdot \nabla \theta}{A}) +\gamma  A \epsilon \sin \theta (\rho_0+A \epsilon \cos \theta -\rho) +\gamma \Omega\label{eq17},\\
& \partial_t A= \Gamma (\nabla^2 A -A (\nabla \theta)^2 )+ 2\chi A (\rho/\rho_0-A^2),\label{eq18}
\end{align}
This gives a minimal model for expansion and contraction patterns of cells observed in the experiments. We note that in our model, the phase is associated with the prefered density of the cells, and one should not compare the absolute value of the phase in the experiments and in the simulations. 
\subsection{Fit to the experimental lengthscale}\label{fitsec}
To explore any correlation between the spatial and temporal patterns, we use the non-monotonic behavior of spatial correlation of velocity as a function of density, derived from experiments and also observed in other studies \cite{angelini2011glass,garcia2015physics}, into the dynamics of the complex field in Eqs. \eqref{eq2} and \eqref{eq3} and measure the average period of the oscillations. First, we non-dimensionalize the equations using the time scale $2 \pi/(\gamma \Omega)$ and the simulation grid lengthscale $\Delta x$. We have
\begin{align}
&\partial_{t'} \rho =  \alpha^{'} \nabla^2( \rho -\rho_0-\epsilon A \cos \theta),\label{eq19}\\
&\partial_{t'} \theta = \Gamma^{'} (\nabla^2 \theta +2 \frac{\nabla A \cdot \nabla \theta}{A}) +\gamma^{'}  A \epsilon \sin \theta (\rho_0+A \epsilon \cos \theta -\rho) +1\label{eq20},\\
& \partial_{t'} A= \Gamma^{'} (\nabla^2 A -A (\nabla \theta)^2 )+ 2\chi^{'} A (\rho/\rho_0-A^2),\label{eq21}
\end{align}
where $\alpha'=2 \pi\alpha/(\Delta x^2 \gamma \Omega)$, $\gamma^{'}=2 \pi/\Omega$, $\Gamma'=2 \pi\Gamma/(\Delta x^2 \gamma \Omega)$, $\chi^{'}=2 \pi \chi/(\gamma \Omega)$.
The coefficient $\Gamma'$  governs the pattern lengthscale in the phase. We thus input $\Gamma^{'}$ using the velocity correlation length from experiments (Fig. \ref{fig:2}g). In particular, we fit a second order polynomial to the square of the velocity correlation function as a function of density, for densities below and above the transition, and use the fits to set the functional form of $\Gamma'$ as a function of density $\rho_0$. We find that the fits $\Gamma(\rho_0) = \Gamma_0(3.9\rho_0^2 - 4.7\rho_0 + 2.6)$ and $\Gamma(\rho_0) = 2.6 \Gamma_0(1 - 0.3 \rho_0)$ resemble the experimental lengthscale squared, with a good $R^2$ value of 0.95, for densities below and above the transition, respectively.

We next solve Eqs. \eqref{eq19}-\eqref{eq21} numerically and find the average phase spatial correlation length, average oscillation period, and average defect density as a function of cell density, as shown in Fig. \ref{figsim1}c-e. The simulations confirm that the temporal and spatial patterns are correlated and that we are able to reproduce the nonmonotonic trends in the temporal period of the phase and phase defects observed in experiments.
\renewcommand{\thefigure}{S\arabic{figure}}
\setcounter{figure}{0}  

\subsection*{Supplementary Figures}
\begin{figure}[!ht]
\centering
  \includegraphics[scale=0.6]{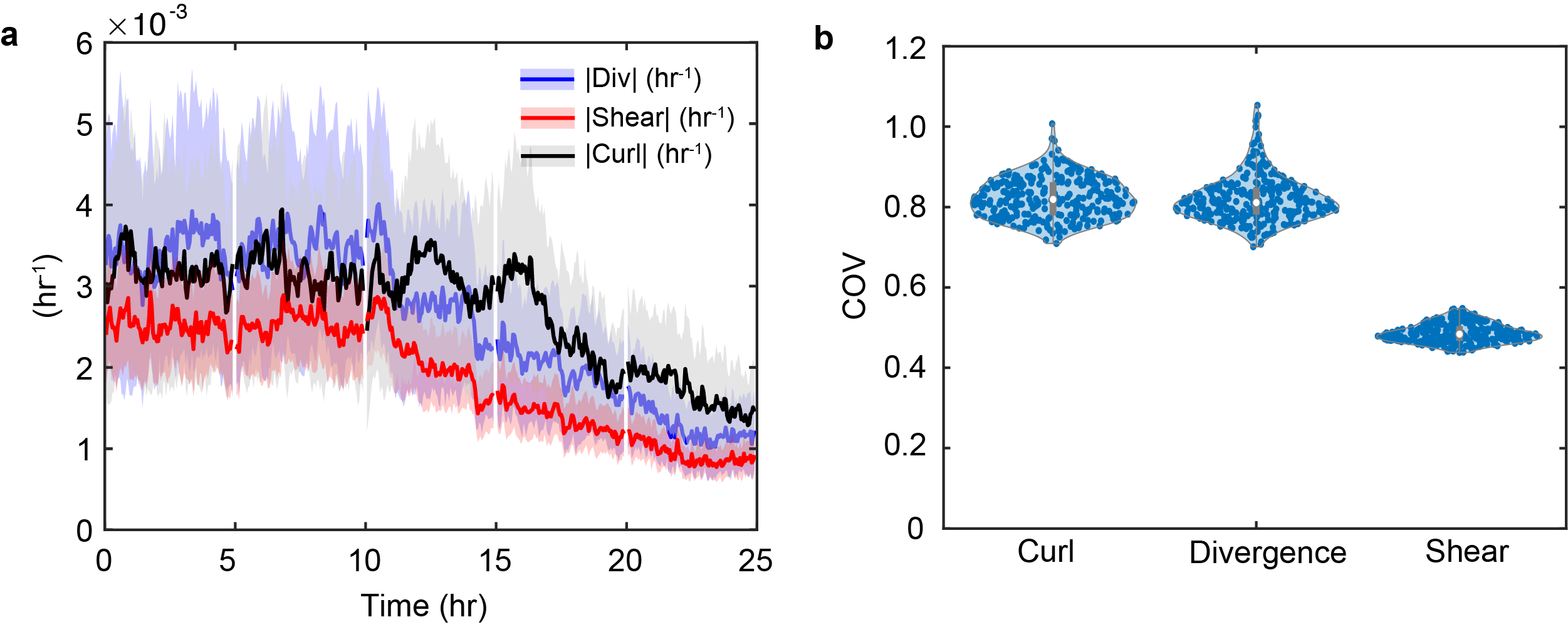}
  \caption{\textbf{Magnitude and Coefficient of Variation (COV) for divergence, curl, and shear} \textbf{a}, Magnitude of divergence, curl, and shear, for the data shown in Fig. 2. \textbf{b}, The corresponding coefficient of variation (COV) for panel \textbf{a}.} 
  \label{fig:S7}
\end{figure}

\begin{figure}[!ht]
\centering
  \includegraphics[scale=0.7]{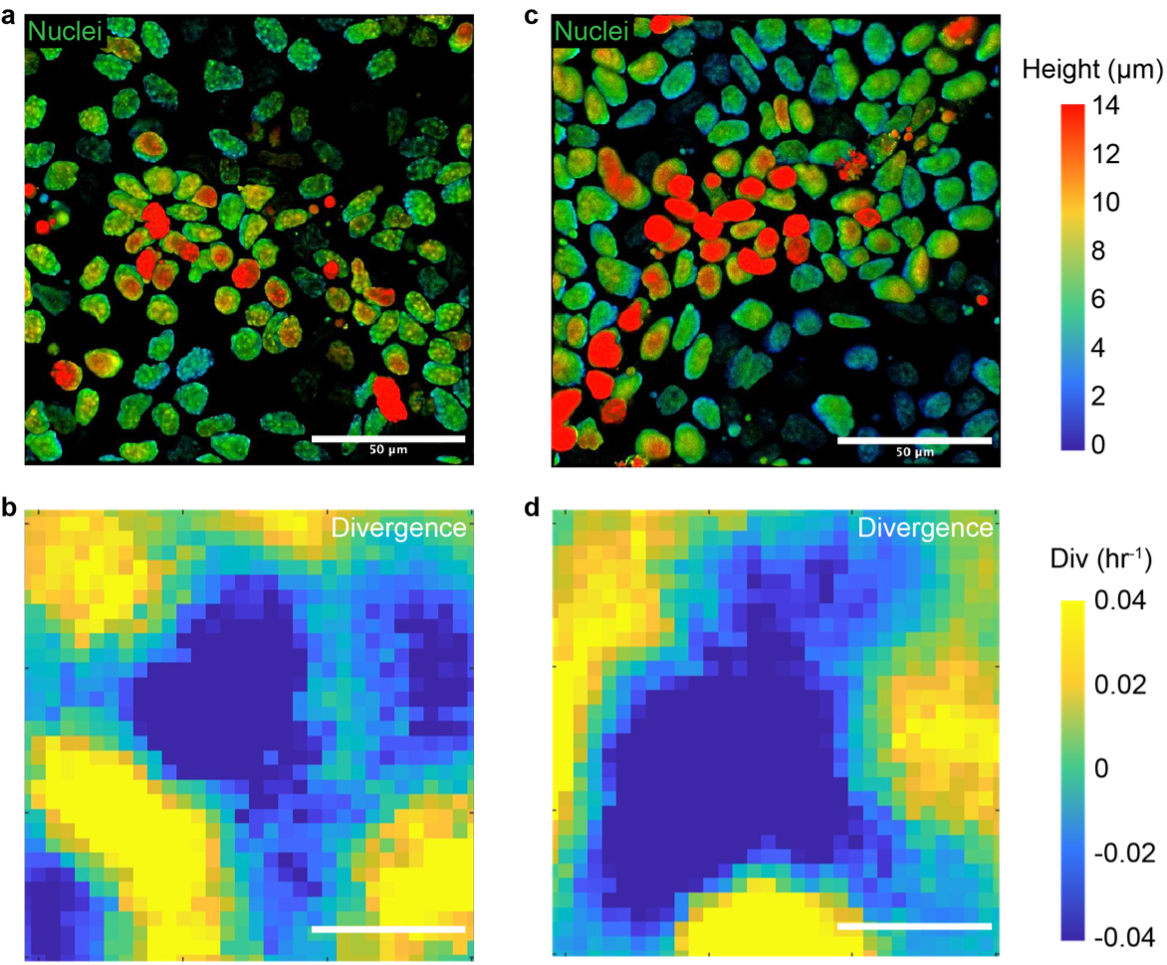}
  \caption{\textbf{Cell height and divergence pattern.} \textbf{a, c}, Cell nuclei height within a small region of the epithelium. \textbf{b, d}, The corresponding divergence patterns for \textbf{a} and \textbf{c}. Scale bars: 50\(\mu m\)} 
  \label{fig:S6}
\end{figure}

\begin{figure}[!ht]
\centering
  \includegraphics[scale=0.9]{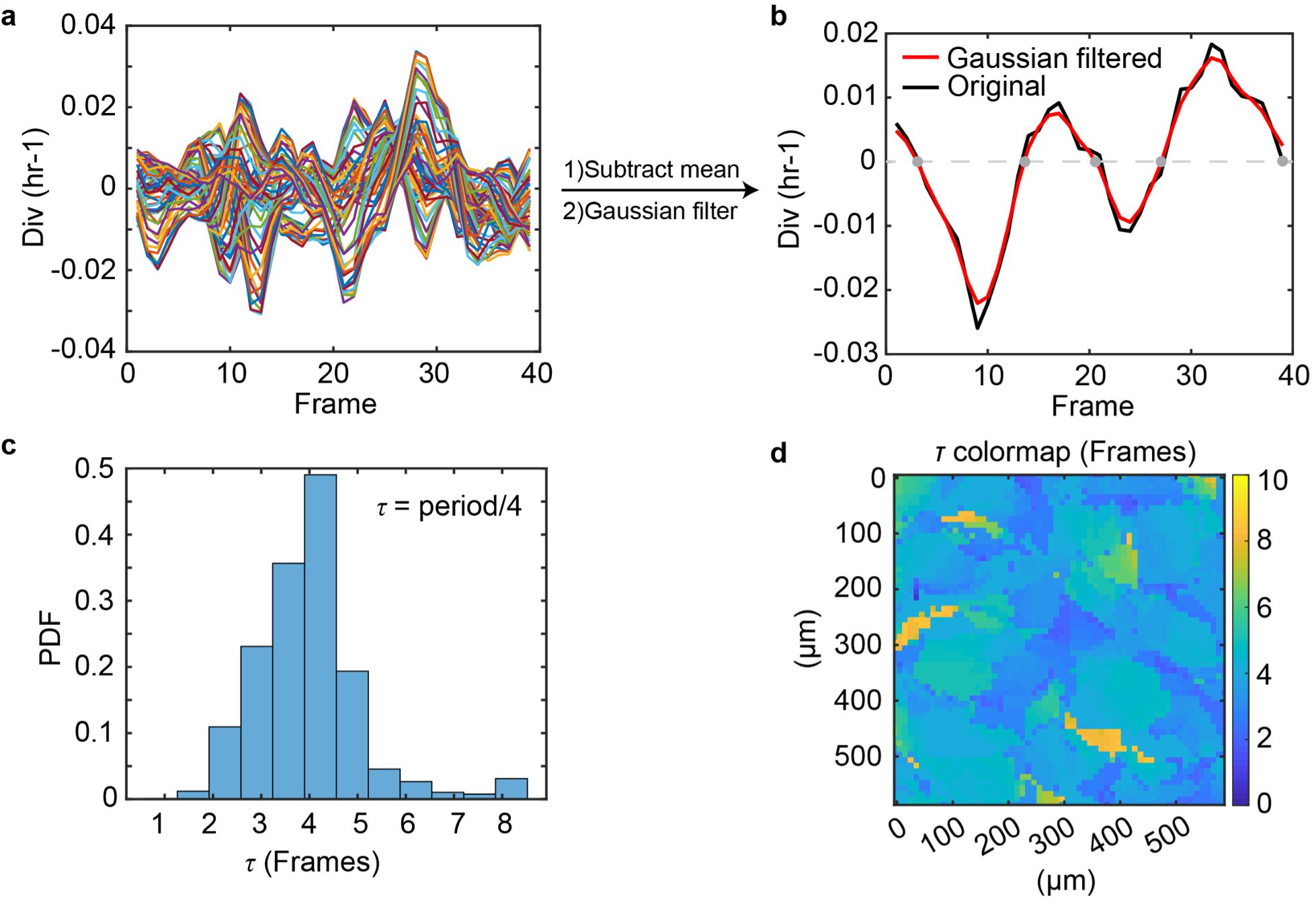}
  \caption{\textbf{Divergence post-processing and period calculation. a}, Divergence variation with time for a series of randomly selected pixels. \textbf{b}, Comparison between divergence \(Div\) with Gaussian filtered divergence \(Div^*\) shows a good agreement. \textbf{c}, The distribution of the quarter period \(\tau\). \textbf{d}, The spatial distribution of the quarter period \(\tau\).}
  \label{fig:S1}
\end{figure}

\begin{figure}[!ht]
\centering
  \includegraphics[scale=0.9]{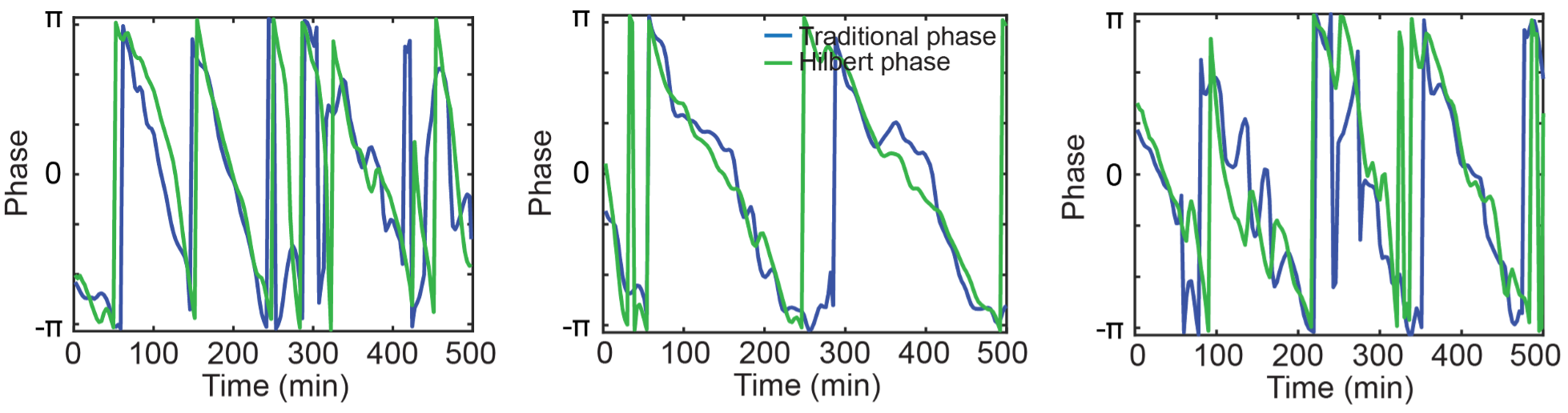}
  \caption{\textbf{Comparision between traditional phase calculation method and Hilbert phase.} Three representative examples show that the traditional phase calculation method agrees well with the Hilbert phase. }
  \label{fig:S2}
\end{figure}

\begin{figure}[!ht]
\centering
  \includegraphics[scale=0.9]{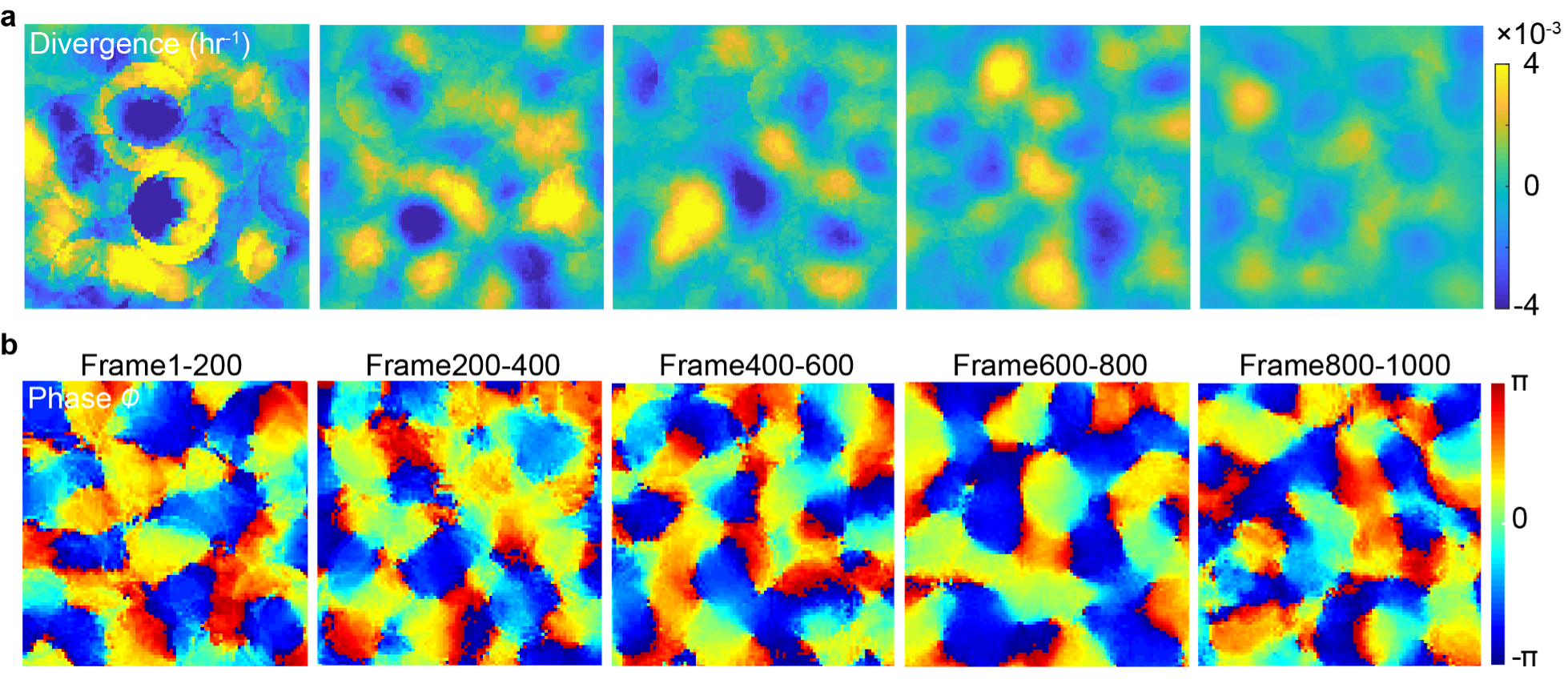}
  \caption{\textbf{Divergence pattern and phase for MDCK cells with increasing density. a}, Divergence colormap as cell number density increases. \textbf{b}, Phase colormap as cell number density increases. }
  \label{fig:S3}
\end{figure}
\begin{figure}[!ht]
\centering
  \includegraphics[scale=0.7]{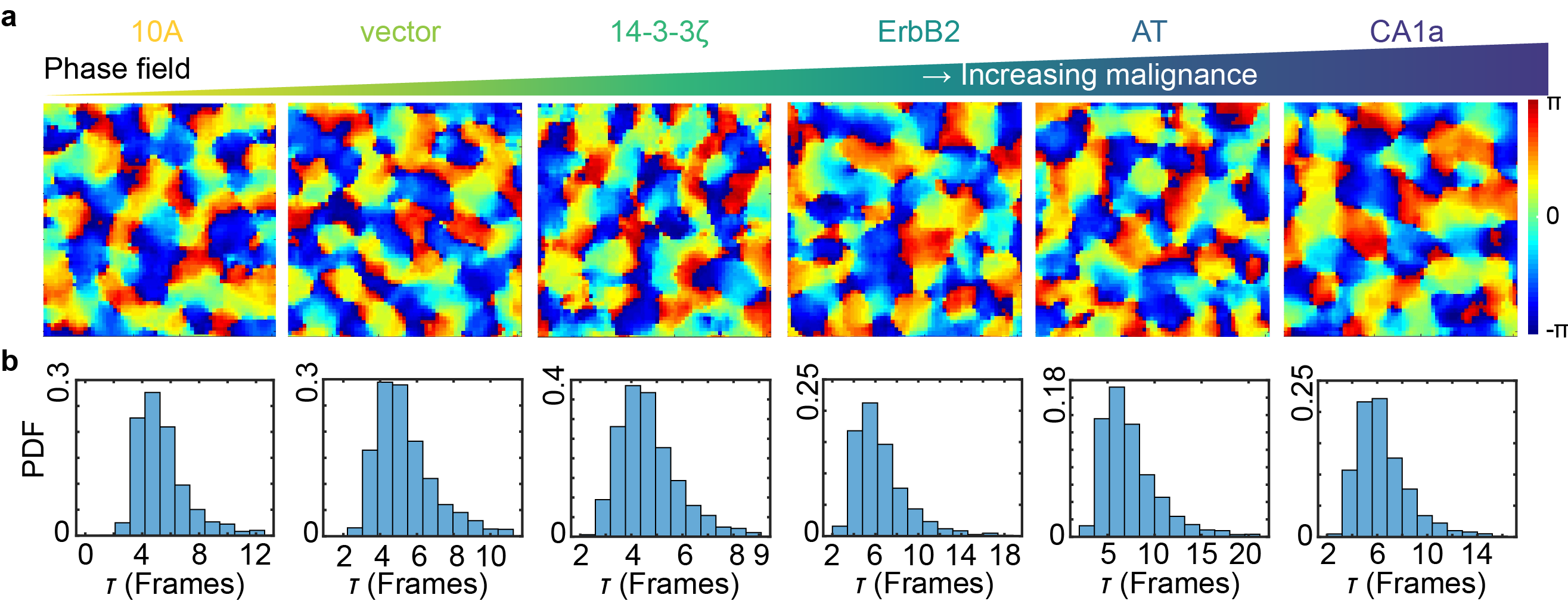}
  \caption{\textbf{Phase pattern for breast cancer cells} (10A, vector, 14-3-3\(\zeta\), ErbB2, AT, CA1a) with increasing malignance levels. \textbf{a}, Phase \(\phi\)  spatial patterns. \textbf{b}, The quarter period \(\tau\) distributions.}
  \label{fig:S4}
\end{figure}

\begin{figure}[!ht]
\centering
  \includegraphics[scale=0.9]{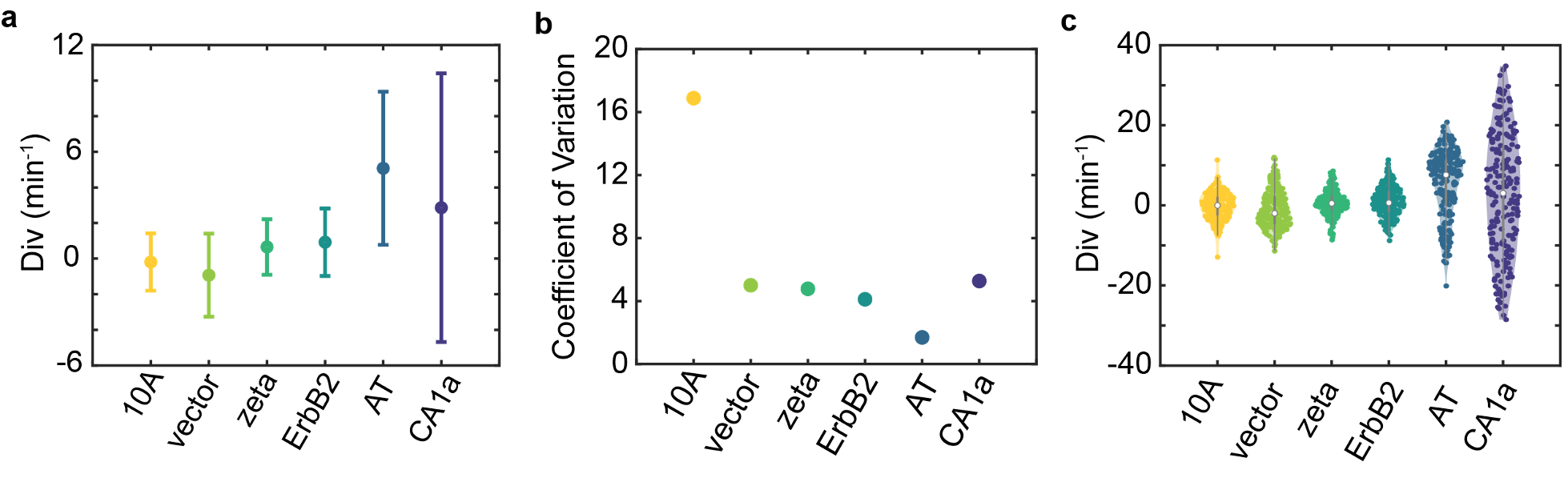}
  \caption{\textbf{Divergence statistics of breast cancer cells with increasing malignancy. a}, Average and standard deviation of divergence for different cell lines. \textbf{b}, Coefficient of variation for divergence defined as the ratio of standard deviation and mean. \textbf{c}, Violin plot for divergence.}
  \label{fig:S5}
\end{figure}

\begin{figure}[!ht]
\centering
  \includegraphics[scale=0.8]{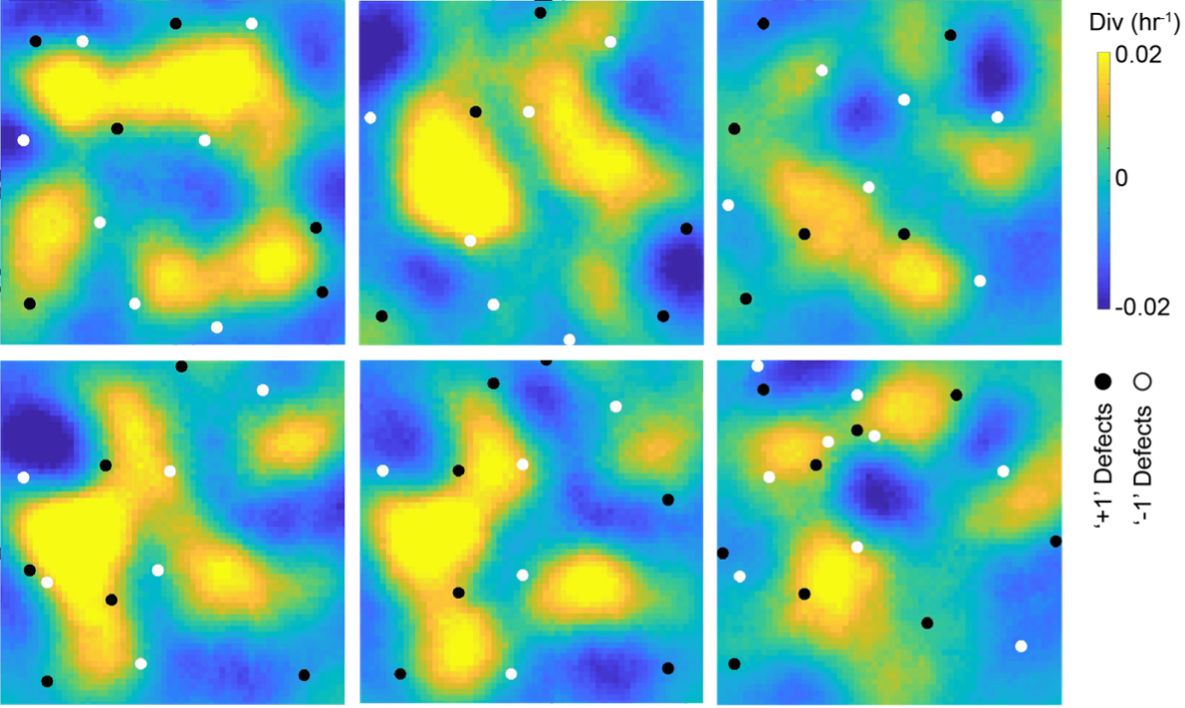}
  \caption{\textbf{Overlay of phase defects with divergence pattern.} Defects appear at the boundary between sources and sinks.}
  \label{fig:S8}
\end{figure}

\begin{figure}[!ht]
\centering
  \includegraphics[scale=0.4]{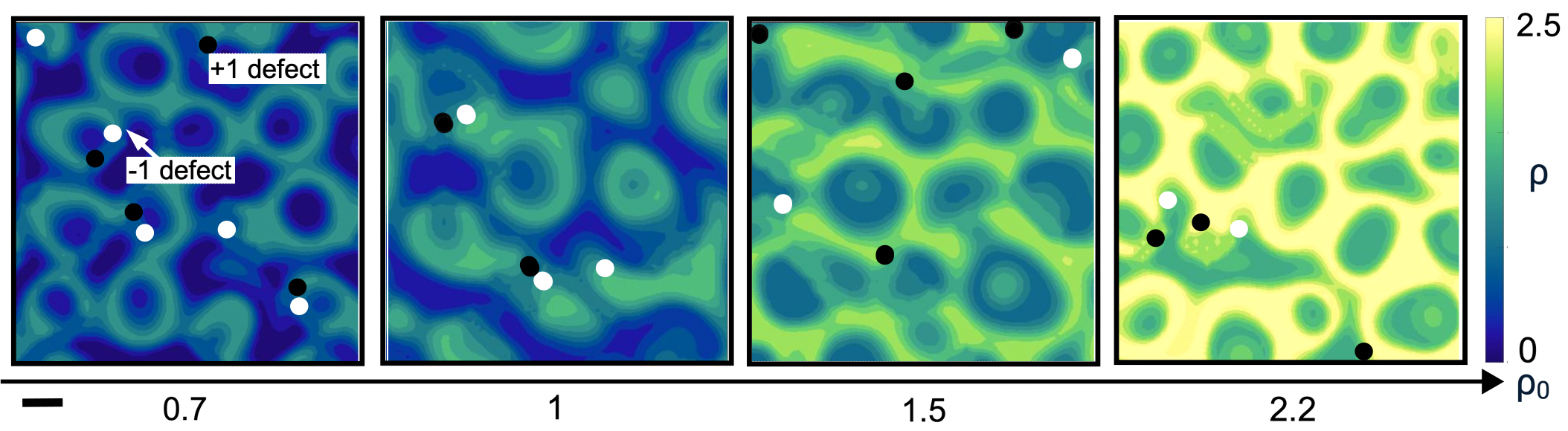}
  \caption{\textbf{Density heatmap overlayed with $+1$ (in black) and $-1$ (in white) defects.} Schematic of density heatmap as average density $\rho_0$ increases. The scale bar represents 15 simulation grid units. }
  \label{fig:S9}
\end{figure}
\end{document}